%% file: master.tex
\documentclass[%
footinbib,
superscriptaddress,
twocolumn,
amsmath,amssymb,
aps,
]{revtex4-1}

\usepackage[utf8]{inputenc}
\usepackage[T1]{fontenc}

\usepackage{graphicx}% Include figure files
\usepackage{dcolumn}% Align table columns on decimal point
\usepackage{bm}% bold math

\usepackage{xcolor}		% For coloring math
\usepackage[english]{babel} % to solve some babel errors

\begin{document}

\preprint{APS/123-QED}

\title{Properties of equilibria and glassy phases of the random Lotka-Volterra model \\
with demographic noise}% Force line breaks with \\

\author{Ada Altieri}
\affiliation{Laboratoire de Physique de l’École normale supérieure, ENS, Université PSL, CNRS, Sorbonne Université, Université de Paris F-75005 Paris, France}

\author{Félix Roy}
\affiliation{Institut de physique théorique, Université Paris Saclay, CEA, CNRS, F-91191 Gif-sur-Yvette, France}
\affiliation{Laboratoire de Physique de l’École normale supérieure, ENS, Université PSL, CNRS, Sorbonne Université, Université de Paris F-75005 Paris, France}

\author{Chiara Cammarota}
\affiliation{Department of Mathematics, King's College London, Strand London WC2R 2LS, United Kingdom}
\affiliation{Dip. Fisica, Università "Sapienza",
Piazzale A. Moro 2, I-00185, Rome, Italy}

\author{Giulio Biroli}
\affiliation{Laboratoire de Physique de l’École normale supérieure, ENS, Université PSL, CNRS, Sorbonne Université, Université de Paris F-75005 Paris, France}

\begin{abstract}
In this letter we study a reference model in theoretical ecology, the disordered Lotka-Volterra model for ecological communities, in the presence of finite demographic noise. Our theoretical analysis, which takes advantage of a mapping to an equilibrium disordered system, 
proves that for sufficiently heterogeneous interactions and low demographic noise the system displays a multiple equilibria phase, which we fully characterize. In particular, we show that in this phase the number of stable equilibria is exponential in the number of species. Upon further decreasing the demographic noise, we unveil a \emph{Gardner} transition to a marginally stable phase, similar to that observed in jamming of amorphous materials. We confirm and complement our analytical results by numerical simulations. 
Furthermore, we extend their relevance by showing that they hold for others interacting random dynamical systems, such as the Random Replicant Model. Finally, we discuss their extension to the case of asymmetric couplings.  
\end{abstract}

\maketitle

\emph{Introduction} --
Lotka-Volterra equations describing the dynamics of interacting species are a key ingredient for theoretical studies in ecology,  genetics, evolution and economy \cite{May2007theoretical, Faust2012,Bucci2014,Goodwin2003, Kessler2015, Maynard2020}. 
Cases in which the number of species is very large are becoming of general interest in disparate fields, such as in ecology and biology, \emph{e.g.} for bacteria communities \cite{Chandler2011, Lloyd2017}, and economy where many agents trade and interact simultaneously both in financial markets and in complex economic systems \cite{risk2012oecd,thurner2011systemic}.\\    
%In recent years there has been an increasing focus on the case in which the number of species is {\it large}. In many situations of current interest in ecology, such as bacteria communities, this is indeed the case \cite{Chandler2011, Lloyd2017};  analogously in economy, where many agents trade and interact simultaneously both in financial markets and in complex economic systems.\\  
The theoretical framework used in the past for a small number of species is mainly based on the theory of dynamical systems \cite{Barreira2013low, Macarthur1970, Tilman1982, Ruan2006, vano2006, van2015}. When the number of ordinary differential equations associated with the Lolta-Volterra (LV) model becomes very large, {\it i.e.} for many species, methods based on statistical physics become ideally suited. Indeed, several authors have recently investigated different aspects of community ecology, such as properties of equilibria, endogeneous dynamical fluctuations, biodiversity, using ideas and concepts rooted in statistical physics of disordered systems  \cite{Kessler2015, Fisher2014, Servan2018, Bunin2017, Biroli2018_eco, Tikhonov2017, Altieri2019,pearce2020stabilization,Roy2020, Marsland2020,sidhom2020ecological,dalmedigos2020dynamical}. Similar investigations have been also performed for economic systems \cite{moran2019may}. The complexity of dealing with a large number of interacting species can actually become a welcome new ingredient both conceptually and methodologically. In fact, different collective behaviours
can emerge. As it happens in physics, such {\it phases} are not tied to the specific model they come from, hence they can be characterized and characterize systems in a generic way \footnote{Along a similar vein, models of liquids and crystals that are used in physics are disparate,   approximate and often inaccurate with respect to real system. Yet, the properties of the phases that arise from their studies provide a precise and quantitative description of the phases found in nature.}. From this perspective, it is natural to ask which kind of different collective behaviors arise from LV models in the limit of many interacting species and what are their main properties \cite{Bunin2017, Biroli2018_eco}. These questions, which have started to attract a lot of attention recently, tie in with the analysis of the properties of equilibria \cite{Fyodorov2016, Fyodorov2018,Fyodorov2020}. \\
Here we focus on the disordered Lotka-Volterra model of many interacting species, which is a representative model of well-mixed community ecology \cite{Barbier2018}, and can be mapped or related to models used in evolutionary game theory and for economic systems \cite{Diederich1989replicators, Galla2013, Sanders2018,Solomon2000, Moran2019}. We consider the case of symmetric interactions and small immigration and work out the phase diagram as a function of the degree of heterogeneity in the interactions and of the strength of the demographic noise. Compared to previous works \cite{Kessler2015,Biscari1995,Bunin2017,Biroli2018_eco} adding demographic noise not only allows us to obtain a
more general picture, but also to fully characterize the phases and connect their properties to the ones of equilibria. In particular, we shall show 
that the number of stable equilibria in the LV model is exponential in the system size and their organization in configuration space follows general principles found for models of mean-field spin-glasses. Our findings, which are obtained for symmetric interactions, provide a useful starting point to analyze the non-symmetric case, as we shall demonstrate by drawing general conclusions on properties of equilibria in the case of small asymmetry. 

\begin{figure}[ht]
\includegraphics[width  = \linewidth]{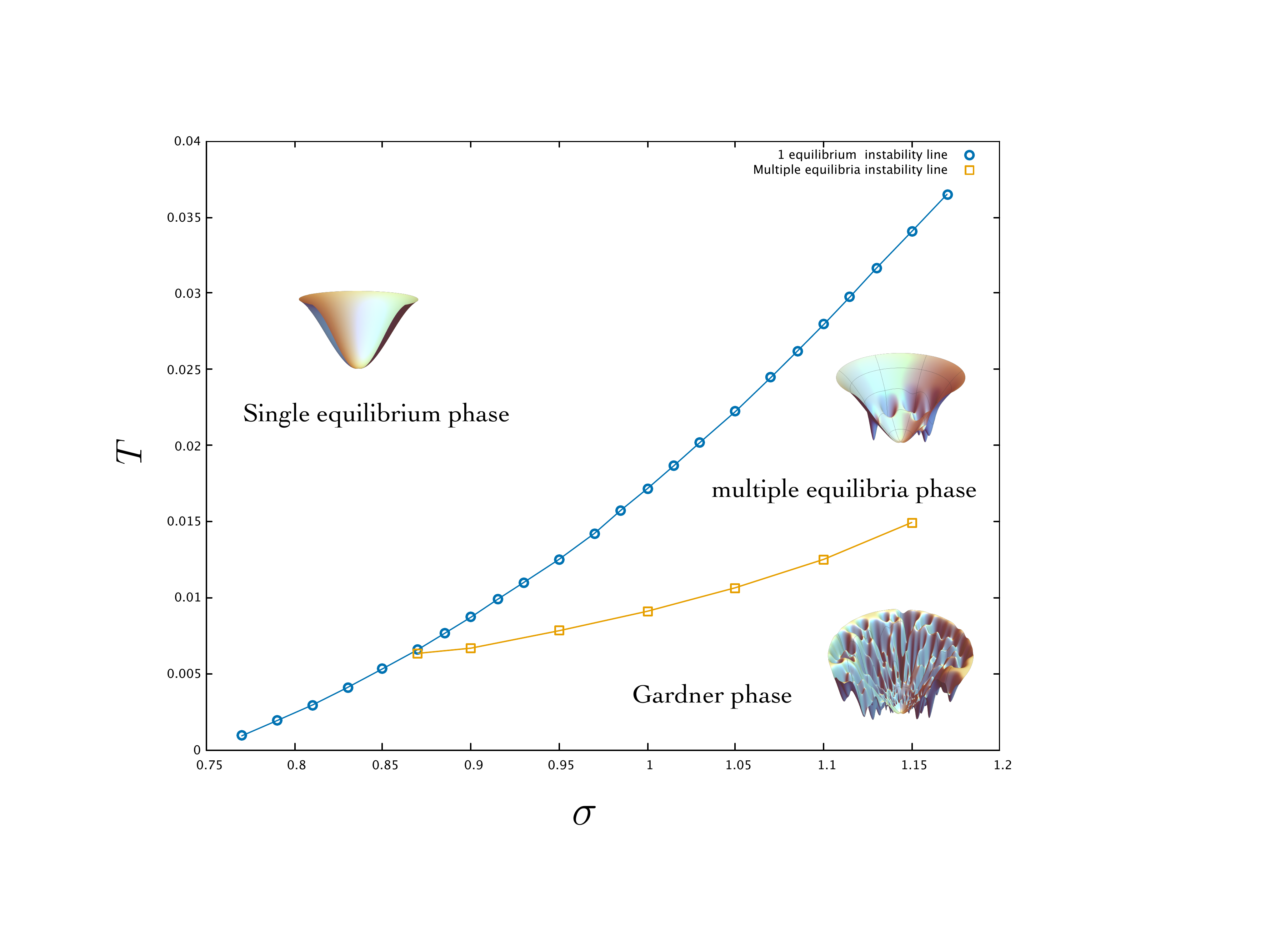}
\caption{Phase diagram showing the strength of the demographic noise, $T$, as a function of the degree of heterogeneity, $\sigma$, at fixed $\mu=10$ and selected value of the cutoff, $N_c=10^{-2}$. Upon decreasing the noise three different phases can be detected: i) a single equilibrium phase; ii) a multiple equilibria regime between the light blue and the orange lines; iii) a Gardner phase, which turns out to be characterized by a hierarchical organization of the equilibria in the free-energy landscape.}
\label{PD_log}
\end{figure}

Henceforth we focus on the disordered Lotka-Volterra model for ecological communities \cite{Kessler2015,Bunin2017} defined by the equations: 
\begin{equation}
    \frac{d N_i}{dt}= N_i \left[1-N_i -\sum_{j, (j \neq i)} \alpha_{ij} N_j \right] +\eta_i(t)
    \label{dynamical_eqT}
\end{equation}
where $N_i(t)$ is the relative abundance of species $i$ at time $t$ ($i=1,\dots S)$, and $\eta_i(t)$ is a Gaussian noise with zero mean and covariance $\langle \eta_i(t)\eta_j(t')\rangle=2 TN_i(t)\delta_{ij} \delta(t-t')$ (we follow Ito's convention). 
This noise term allows us to include the effect of demographic noise in a continuous setting \cite{Domokos2004discrete, Rogers2012, weissmann_simulation_2018}; the larger is the global population the smaller is the strength, $T$, of the demographic noise.  Immigration from the mainland is modeled by a reflecting wall for the dynamics at $N_i=\lambda$, since this is more practical for simulations than the usual way of adding a $\lambda$ in the RHS of Eq. (\ref{dynamical_eqT}) (see the Appendix for more details).  
%We have found that the usual way, that consists in adding a $\lambda$ term on the right-hand side of (\ref{dynamical_eq}), is problematic in presence of demographic noise since it induces a lower cut-off on the abundances at a scale $e^{-T/\lambda}$ which can be extremely small for $\lambda \ll T$.  
The elements of the interaction matrix $\alpha_{ij}$ are independent and identically distributed variables such that:
\begin{equation}
\text{mean}[\alpha_{ij}]=\mu/S \hspace{0.5cm} \text{var}[\alpha_{ij}]=\sigma^2/S \ ,  
\end{equation}
%{\color{red}{Nonetheless, these random variables can be drawn from any distribution without long tails as all that matters are their mean and variance.}}
that we consider in the symmetric case with $\alpha_{ij}=\alpha_{ji}$.  As shown in \cite{Biroli2018_eco}, the stochastic process induced by eq. (\ref{dynamical_eqT}) admits an {equilibrium}-like stationary Boltzmann distribution:
\begin{equation}
    P(\{ N_i\})=\exp\left(-\frac{H(\{ N_i\}}{T}\right)
\end{equation}
where 
\begin{equation}
\begin{split}
    H= & {{-}} \sum_i  \left(N_i-\frac{N_i^2}{2}\right)+\sum_{i<j}\alpha_{ij}N_iN_j+\\
    & + \sum_i [T\ln N_i + \ln \theta(N_i-\lambda)]
    \label{Hamiltonian_0}
    \end{split}
\end{equation}
The before-last term is due to the demographic noise and the last one to the reflecting wall, which leads to a lower-immigration cut-off, at $N_i=\lambda$ ($\theta(x)$ is the Heaviside function). By taking advantage of this mapping to an equilibrium statistical mechanics problem and by using theoretical methods developed for disordered systems, we obtain the properties of the stationary states and the equilibria of the LV-model from the analysis of the equilibrium states and the local minima of the energy function $H$. Our theoretical framework is standard and based on the replica method \cite{MPV}; the computation is described in full details in the Appendix. Here, we present directly the results. \\
Among the most important ones is the existence of three distinct phases for the LV-model in presence of demographic noise and small but non-zero immigration, as shown in Fig. \ref{PD_log} (we focus on $N_c=10^{-2}$, similar results are obtained for smaller values of $N_c$). 
We find no sensitive dependence on the average interaction parameter, so the phase diagram has been obtained at fixed value $\mu=10$. More details will follow in the Appendix. 

For large enough demographic noise (corresponding to high-temperature) we find that there is a single equilibrium phase, \emph{i.e.} the noise is so strong that the interactions within species do not play an important role: for any initial condition the system relaxes toward a unique dynamically fluctuating stationary state. When the strength of the demographic noise decreases, multiple states emerge. We can study this transition by analyzing the stability of the thermodynamic high-temperature phase. This is performed by analyzing its free-energy Hessian matrix $\mathcal H$. The point at which the lowest eigenvalue of $\mathcal H$ reaches zero signals the limit of stability of the high-temperature phase and the emergence of multiple equilibria. 

Within the replica method that we used here, this corresponds to the breaking of replica symmetry and to the requirement of having a zero replicon eigenvalue. As explained in the Appendix, this leads to an equation for the transition line corresponding to the blue curve in Fig.1:
\begin{equation}
\lambda_\text{R}=(\beta  \sigma)^2 \left[ 1-(\beta \sigma)^2 \overline{ \left(\langle N_i^2 \rangle - \langle N_i \rangle^2 \right)^2} \right]=0 \ ,
\end{equation}
where $\beta=1/T$. The average $\langle \cdot \rangle$ is the thermodynamics average taken over the effective Hamiltonian (\ref{Hamiltonian_0}), while $\overline \cdot$ denotes the average over the quenched disorder associated to the random interactions ($i$ is a dummy index since statistically all species are equivalent after average over the interactions). Physically, the condition above 
can be shown to correspond to a diverging response function \cite{Bunin2017,Biroli2018_eco}, and is a signature of the system being at the edge of stability, namely at a \emph{critical point} in the parameter space.

%The phase diagram in absence of demographic noise and with infinitesimally small immigration, $\lambda \rightarrow 0^{+}$, has been already obtained in \cite{Biroli2018_eco}.

%Our findings agree with 
%In this work we provide a much more detailed and complete framework, integrating both the presence of demographic noise and the contribution of the immigration via a cut-off on the species abundances, $N_c$.

\begin{figure}
    \centering
    \includegraphics[scale=0.21]{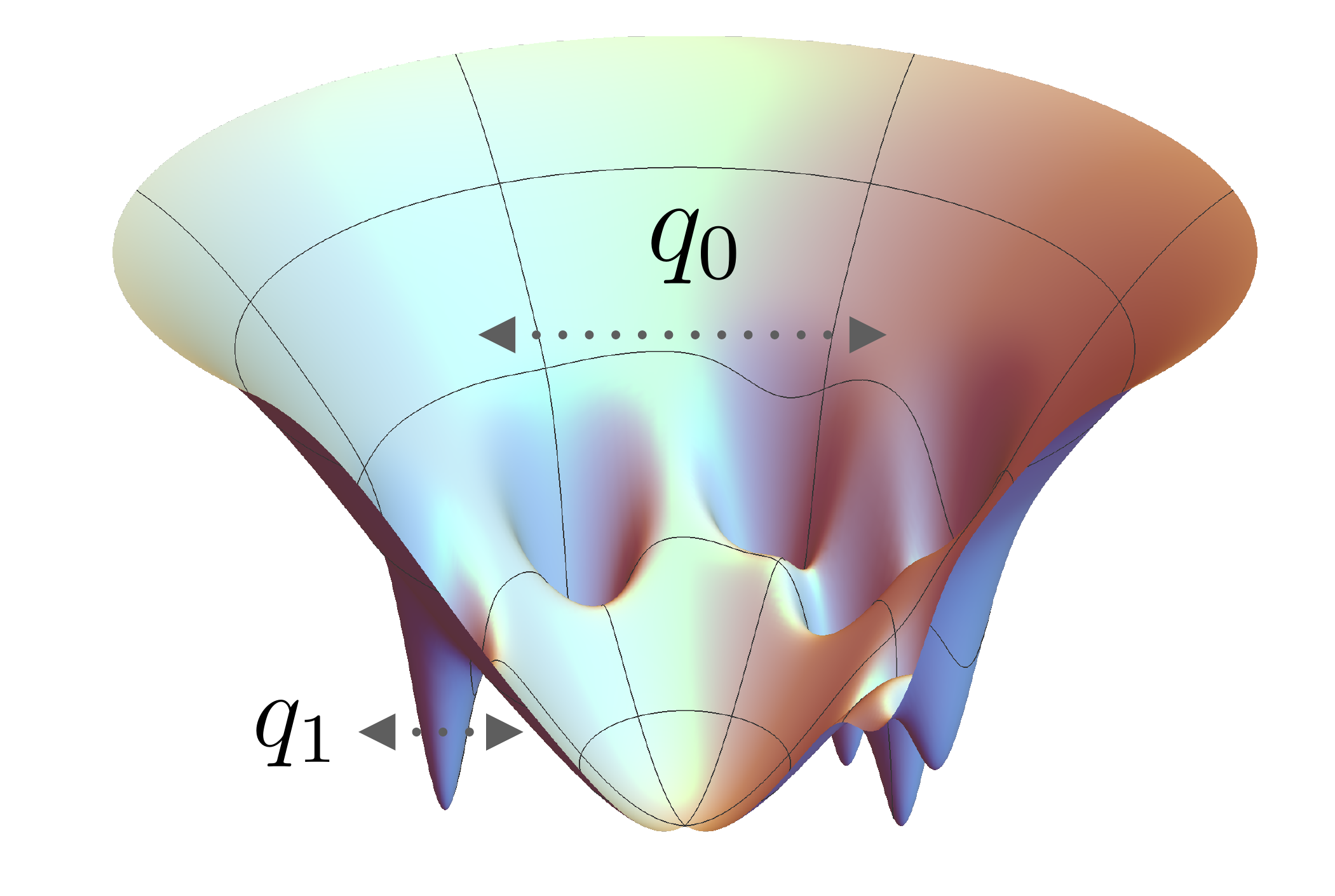}
    \caption{Zoom on a pictorial landscape. The parameters $q_0$ and $q_1$ denote the size of the largest and the innermost basins respectively within the two-level structure of the $1$RSB phase.}
    \label{landscapes_1_full}
\end{figure}

Below the blue curve there exist multiple states---which one is reached dynamically depends on the initial condition. Such states correspond to dynamically fluctuating equilibria that are stable to perturbations and that have typically an overlap in configuration space given by 
\begin{equation}
q_0=\frac 1 S \sum_i\langle N_i \rangle_\alpha\langle N_i \rangle_\beta \ ,
\end{equation}
where $\alpha$ and $\beta$ denote the average within two generic states $\alpha$ and $\beta$. One can similarly define the intra-state overlap $q_1=\frac 1 S \sum_i\langle N_i \rangle_\alpha^2$. See Fig. \ref{landscapes_1_full} for a pictorial representation of these two quantities and the organization of equilibria in phase space. This is (in the replica jargon) the so-called one-step replica symmetry breaking phase ($1$RSB). In order to characterize the properties of this phase of the LV-model, we have computed the number of states, and hence of equilibria, using methods developed for structural glasses \cite{Monasson1995}. More specifically, we have computed the complexity $\Sigma$ (see the Appendix for details), which is defined as the logarithm of the number of equilibria with a given free-energy density $f$ normalized by the number of species $S$. 
%We have obtained the general formula:
%\begin{equation}
%\begin{split}
 %   \Sigma= & m^2 \frac{d}{d m} \left( \beta F^{\text{$1$RSB}}\right)= \\
%= & \frac{m^2  \sigma^2 \beta^2}{4}(q_1^2-q_0^2) +\int \mathcal{D}z \; \ln \int \frac{d t_{a_B}}{\sqrt{2 \pi}}e^{-\frac{t_{a_B}}{2}}A(z,t_{a_B})^m \\    
% & - m  \int  \mathcal{D} z \frac{\int \frac{d t_{a_B}}{\sqrt{2 \pi}}e^{-\frac{t_{a_B}}{2}}A(z,t_{a_B})^m \ln A(z,t_{a_B})}{\int \frac{d t_{a_B}}{\sqrt{2 \pi}} A(z,t_{a_B})^m}
%    \end{split}
%\end{equation}
%where $q_1=\frac 1 S \sum_i \langle N_i\rangle^2_\alpha$ and 
% $q_d=\frac 1 S \sum_i \langle N_i^2\rangle_\alpha$ are the overlap and the self-overlap between configurations inside the same state, see Fig. \ref{landscapes_1_full} for a pictorial representation; $m$ is a parameter conjugated to the free-energy $f$, i.e. changing it we scan equilibria with different values of $f$. As shown in the appendix, we can analytically obtain the values of $q_1$ and $q_d$ and we have an explicit expression of the function  $A(z,t_{a_B})$. This allows us to 
This allow us to show that the number of equilibria below the blue line in Fig. \ref{PD_log} is {\it exponential} in $S$, \emph{i.e.} there is a finite complexity $\Sigma$. 

%The function $A(z,t_{a_B})$ is introduced to denote
%\begin{equation}
%A(z,t_{a_B}) \equiv  \int d N e^{-\beta H_\text{$1$RSB}(N,z,t_{a_B})} 
%\end{equation}
%whereas the parameter $m$, that appears also in Eq. (\ref{average_1rsb}), is peculiar to the static replica formalism, giving information on those states which are \emph{marginally stable}, also called threshold states. The critical slowing down of the dynamics and related aging phenomena are a consequence of the flatness of the free energy around these states.

When decreasing further the demographic noise, the heterogeneity in the interactions becomes even more important and a second phase transition takes place. In order to locate it, we repeat exactly the same procedure as for the single equilibrium phase but now within one of the typical states
with a given free-energy $f$ \footnote{we focus on the ones giving the leading contribution to the partition function; considering a different value would just slightly shift the transition line but keeps qualitatively unaltered the conclusions}. The computation is more involved (it corresponds to analyze the stability of the $1$RSB Ansatz) and leads to the condition:
\begin{equation}
\lambda_\text{R}^\text{1rsb}=(\beta \sigma)^2 \left[ 1-(\beta  \sigma)^2 \overline{ \langle \left(\langle N^2 \rangle_\text{$1$r} - \langle N \rangle^2_\text{$1$r} \right)^2 \rangle_\text{$m$-r}} \right] =0
\label{average_1rsb}
\end{equation}
where the two different averages correspond to i) the intra-state average $\langle \cdot \rangle_{\text{$1$r}}$ and the inter-state average, $\langle \cdot \rangle_\text{$m$-r}$. 
All technical details of the calculation will follow in the Appendix.
The critical temperature that results from the equation above leads to the orange line in Fig. \ref{PD_log}. Crossing this line results in a fragmentation of each state into a fractal structure of sub-basins \cite{Charbonneau2014} (see the landscape on the bottom in Fig. \ref{PD_log}): each state becomes a meta-basin that contains many equilibria, all of them marginally stable, i.e. poised at the edge of stability \cite{Biroli2018_eco}, and organized in configuration space in a hierarchical way, as it was discovered for mean-field spin glasses \cite{MPV}.
This phase, which is called \emph{Gardner}, plays an important role in the physics of jamming and amorphous materials \cite{Berthier2019}. Our results unveil its relevance in theoretical ecology by showing that it describes the organization of equilibria in the symmetric disordered LV-model at low enough demographic noise and for highly heterogeneous couplings.

We now present numerical simulation results that confirm and complement our analytical study. We numerically integrate the stochastic equation Eq. (\ref{dynamical_eqT}) using a specifically designed method (see the Appendix for details). The initial abundance 
for each species is drawn independently in $[0,1]$.  

\begin{figure}[ht]
\includegraphics[width  = \linewidth]{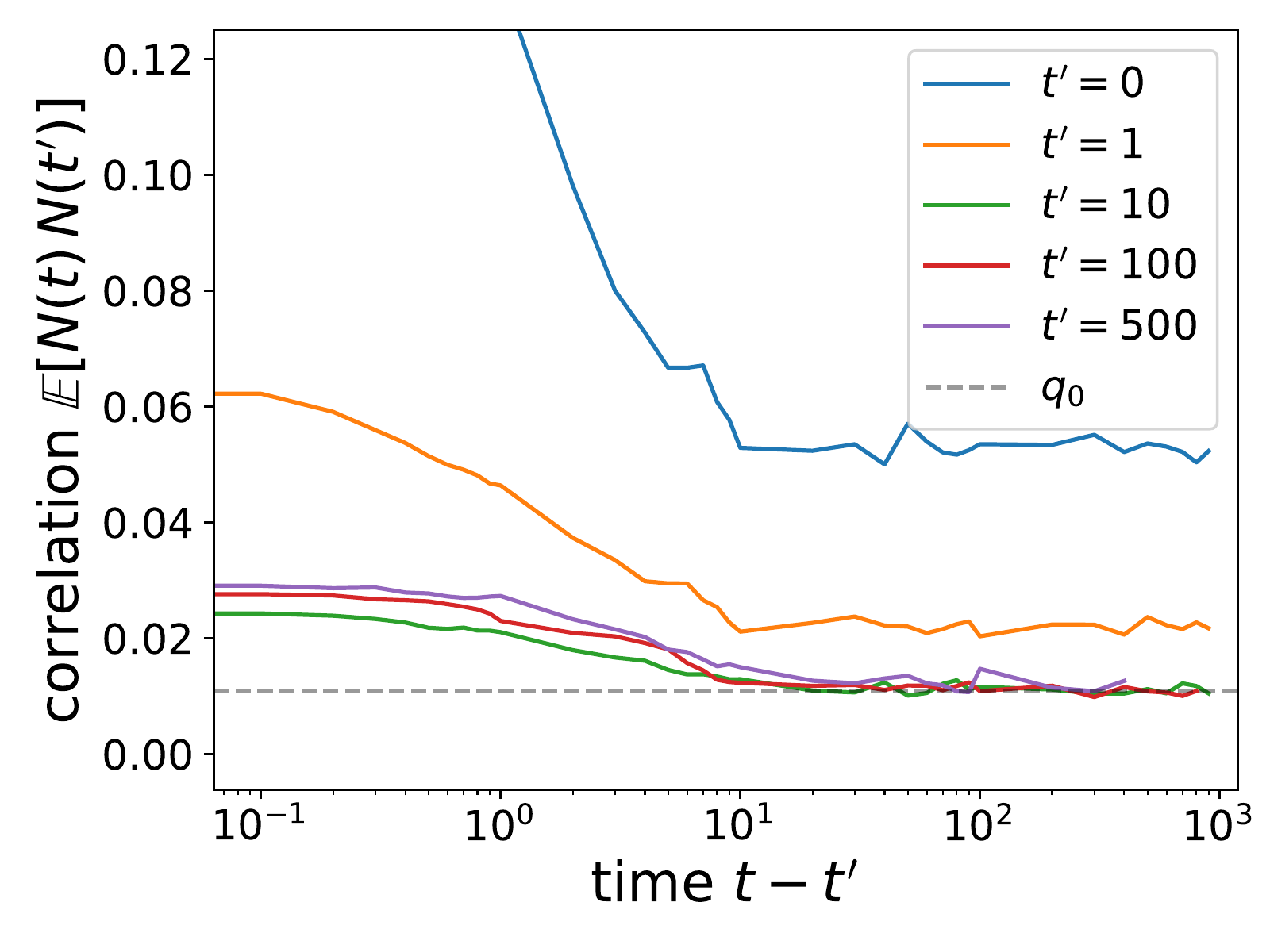}
\caption{Correlation function $C(t,t')$ as a function of $t-t'$ in the one-equilibrium (Replica Symmetric) phase. The curves clearly collapse to the theoretically predicted value, $q_0$, for times $t'>t_{\text{wait}}$.}
\label{correlator_RS}
\end{figure}

\begin{figure}[h]
\includegraphics[width  = \linewidth]{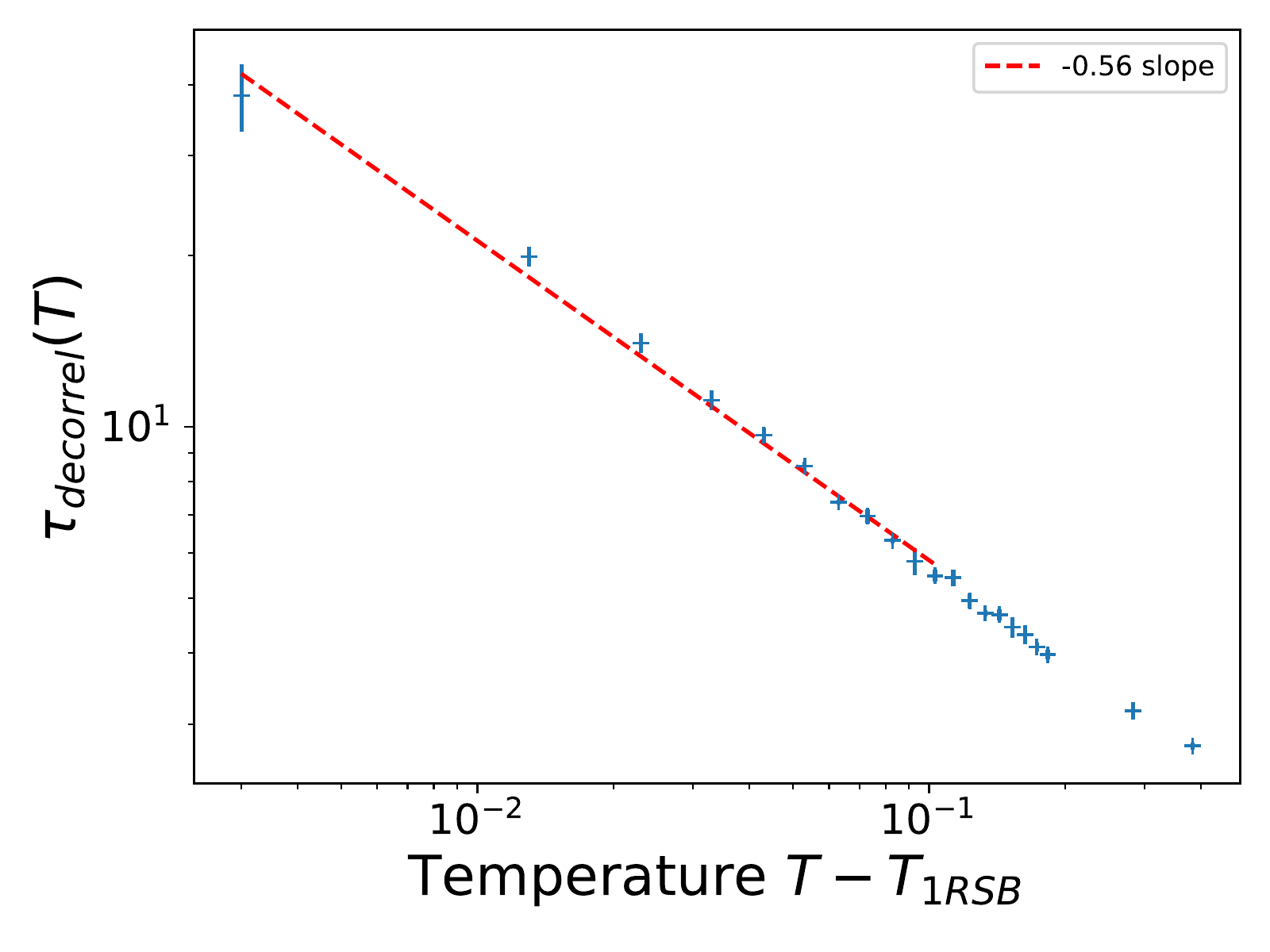}
\caption{Decorrelation time as a function of $(T-T_{\text{1RSB}})$ in logarithmic scale.  The blue points correspond to numerical data, while the dashed red line is a fit. The decay of the decorrelation time in $(T-T_\text{1RSB})$ occurs with an exponent $ \approx -0.5$.}
\label{taucorrelation}
\end{figure}

There are three sources of randomness for a given sample: the interactions, the initial conditions and the demographic noise. In the following, we obtain numerically the average correlation function defined by
\begin{equation}
\mathbb{E}[N(t)N(t')] = \frac{1}{S N_{sample}} \sum_{i=1}^S   \sum_{r=1}^{N_{sample}}  N_i^r(t) N_i^r(t')
\end{equation}
where $\mathbb{E}[X]$ stands for the average over all those sources of randomness.
If the system size is sufficiently large ($S \gg 1$) as well as the sampling set, with $N_\text{sample} \gg 1$, it can be shown that the stochastic process converges in law \cite{benarous_convInLaw}. 
We generically choose $S \sim 500$ and $N_\text{sample} \sim 50$ and eventually verify there is no $(S,N_{sample})$-dependency at this scale. 
We find that in the high-temperature phase 
 a time-translationally invariant (TTI) state is reached after a finite time-scale $t_{wait}$:
\begin{equation}
\forall t \ge t' > t_{wait} \hspace{0.6cm}    \mathbb{E}[N(t)N(t')]=C(t,t')\simeq C(t-t') \ .
\end{equation}
This convergence to a TTI regime is shown in Fig. \ref{correlator_RS}. The long-time limit of $C(t-t')$ is the overlap between two generic configurations belonging to the single equilibrium state: the dashed line in Fig. \ref{correlator_RS} is the analytical prediction for $\lim_{t-t'\rightarrow \infty} C(t-t')$ which is in perfect agreement with the numerics. We have also checked that this agreement holds upon varying $T$ and for other observables, the results are reported in the Appendix (see Fig. \ref{fig:RS_theory}).
From the time-dependence of $C(t-t')$ one can estimate the typical time-scale characterizing dynamical fluctuations within the single equilibrium phase. Formally, we define $\tau_{decorrel}$ by the identity:
\begin{equation}
\frac{C(\tau_{decorrel}) - C(\infty)}{\left( C(0) - C(\infty) \right)} = 0.3 \, 
\end{equation}
In Fig. \ref{taucorrelation}, we plot $\tau_{decorrel}$ as a function of $(T-T_{\text{$1$RSB}})$, where $T_{\text{1RSB}}$ is the critical value of $T$ at which the single equilibrium phase becomes unstable (blue line in Fig. \ref{PD_log}). 
We find that the thermodynamic instability is accompanied by a dynamical transition at which $\tau_{decorrel}$ diverges as a power law with an exponent close to $0.5$, see Fig. \ref{taucorrelation}.

\begin{figure}[hbtp]
\centering
\includegraphics[width  = \linewidth]{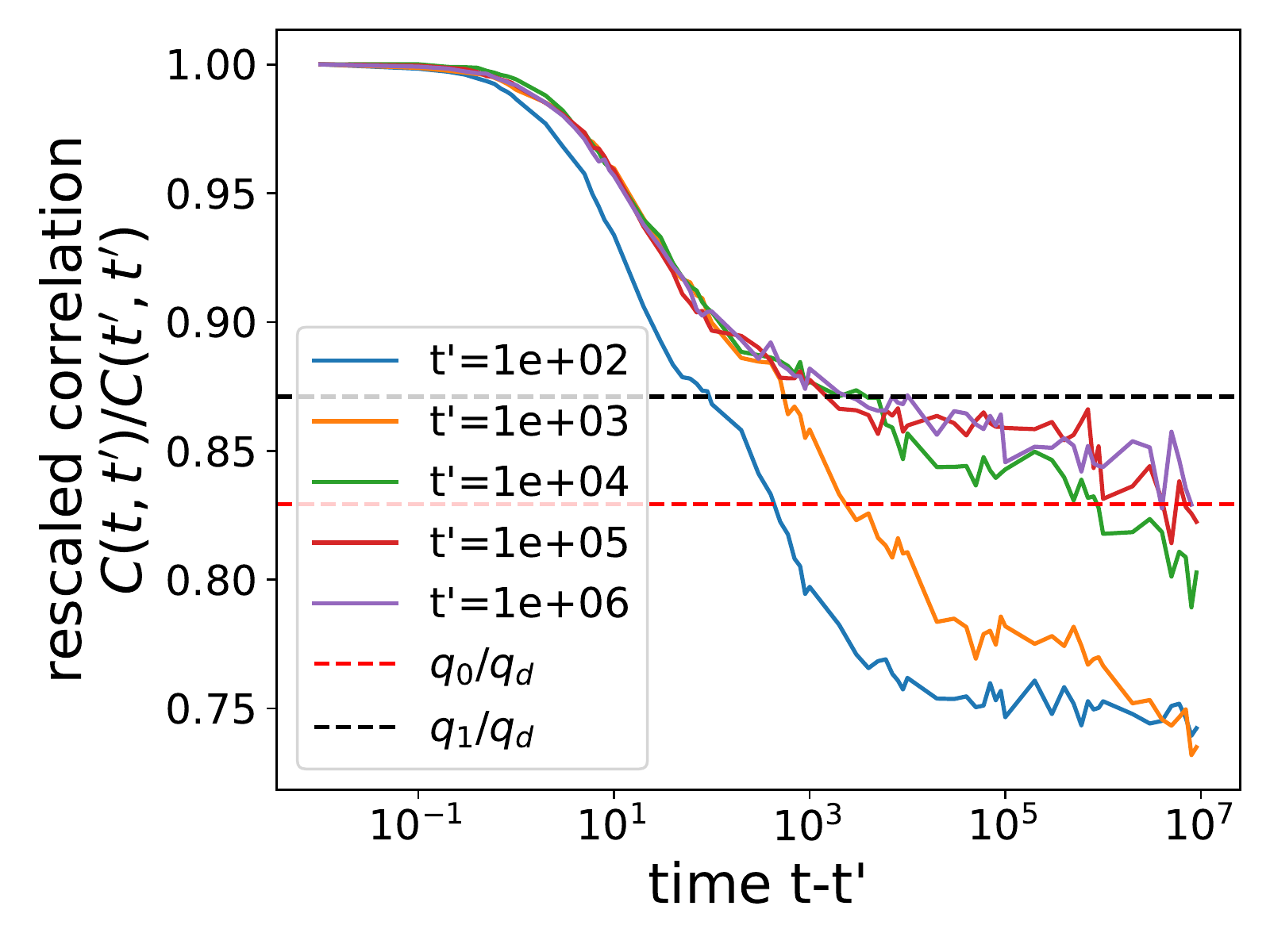}
\caption{Rescaled correlation as a function of $(t-t')$, for different $t'$, showing aging dynamics. The dashed black and red lines correspond respectively to the theoretical predictions for $q_1$ and $q_0$ both rescaled by the analytical prediction for $C(t't')$ (called $q_d$ in the Appendix).}
\label{correlation_RSB}
\end{figure}

For small demographic noise, \emph{i.e} when $T$ is below the blue line of Fig. \ref{PD_log}, previous results on the dynamics of mean-field spin glasses \cite{Sompolinsky1982, Franz1994, Cu-Ku1993} suggest that the LV-model should never reach an equilibrium stationary state, and instead it should display \emph{aging} \cite{Cugliandolo2003, Biroli2005}. 
In fact, one expects that among the very many equilibria the dynamics starting from high-temperature-like initial conditions falls in the basin of attraction of the most numerous and marginally stable equilibria, and display aging behaviour. 
This is indeed what we report in Fig. \ref{correlation_RSB} which shows that the longer is the age of the system, $t'$, the longer it takes to decorrelate. The {landscape} interpretation of this phenomenon is that the system approaches at long times a part of configuration space with many marginally stable equilibria. This leads to aging because the longer is the time, the smaller is the fraction of unstable directions to move, hence the slowing down of the dynamics, but the exploration never stops and eventually the system never settles down in any equilibrium \cite{kurchan1996phase,Cu-Ku-Peliti1997,Parisi2006dynamics}. The two dashed lines in Fig.  \ref{correlation_RSB} correspond to our analytical prediction for the intra-state and the inter-state overlaps of the marginally stable equilibria. The agreement is satisfactory but larger times would be needed to fully confirm it.

Our characterization of the phases and the dynamics of the LV-model has important consequences on related systems, in particular on the so-called random replicant models (RRMs) that consist of an ensemble of replicants evolving according to random interactions. Given their numerous applications in biology, optimization problems \cite{Mende1986, Diederich1989replicators} as well as evolutionary game theory \cite{Smith1982, Sigmund2011}, RRMs still attract great theoretical interest. The RRM, which was introduced in \cite{Diederich1989replicators} and further studied in \cite{Biscari1995}, is remarkably similar to the disordered LV-model we studied. In the case of symmetric interactions, 
one can similarly map the problem onto an equilibrium statistical physics one with the following Hamiltonian:
\begin{equation}
H_{R}=-\sum_{i<j=1}^S J_{ij} x_i x_j - a \sum_{i=1}^S x_i^2    
\end{equation}
where $x_i/S$ is the concentration of the $i$th family in the species pool subject to the global constraint $\sum_{i}x_i=S$ for all $x_i \ge 0$.
The couplings $J_{ij}$ are i.i.d. Gaussian variable with variance $J^2/S$. Provided an appropriate rescaling of the interaction matrix of the two models, we can show that the average interaction term $\mu$ for LV -- standing for a purely competitive environment -- plays the same role as the Lagrange multiplier that is introduced in RRM to enforce the sum of all concentrations to be fixed.
The main differences with respect to Eq. (\ref{Hamiltonian_0}) is the absence of the logarithmic term. Our analysis can be fully extended to the RRM, as we show in the Appendix.  
The main result is that the three phases we found for the LV-model are present also for the RRM, and organized in a phase diagram (see Fig. 8) that is remarkably similar to the one in Fig. \ref{PD_log}. This strengthens the generality of our results, and clarifies the nature of the glassy phase of the RRM that was first investigated in \cite{Biscari1995}.\\
Let us finally discuss how we expect our results to change if the interactions contain a small random asymmetric component. The multiple basins structure associated with the $1$RSB phase should not be affected because its basins correspond to stable stationary states, and a small non-conservative random force should not destabilize them \cite{Berthier2013asym}. On the contrary, the fractal structure and the 
decomposition into sub-basins are expected to be wiped out because of the marginal stability of the equilibria associated with it \cite{Hertz1986, Hertz1987,Fyodorov2020}. In absence of demographic noise, one therefore expects a single equilibrium at small $\sigma$, which 
is replaced by an exponential number of chaotic attractors at large $\sigma$. The demographic noise adds additional dynamical fluctuations to these multiple equilbria and eventually makes them merge in a single equilibrium, thus leading to a phase diagram similar to Fig.1 but only with the blue line and two phases (single and multiple equilibria).    

%The authors  of \cite{Biscari1995} showed there exists a critical value of the control parameter $a$ -- aiming to limit the unbounded growth of one single species --  below which the system develops non-ergodic features and symmetry breaking effects. This emerging criticality precisely corresponds to the critical value of the heterogeneity parameter $\sigma$ in the Lotka-Volterra model without demographic noise, \emph{i.e.} $\sigma_c=1/\sqrt{2}$ \cite{Biroli2018_eco}.

%We will devote part of the Appendix to discuss possible links to our model and derive the corresponding phase diagram without the logarithmic contribution. Even in this second case, we highlight the presence of an amorphous (marginal) phase in the limit of zero demographic noise, which was never revealed before.

In conclusion, we have unveiled a complex and rich structure for the organization of equilibria in a central model for ecological communities. 
Our results, supported by dynamic simulations, highlight the relevance of multiple equilibria phases for the dynamics of many strongly interacting species. Moreover, our findings clarify the glassy nature of the equilibria previously studied in \cite{Kessler2015, Bunin2017, Biroli2018_eco, Altieri2019, Altieri2019book}. As we have shown, our results carry out to more general contexts, in particular to models originating from evolutionary game theory. We expect that the collective dynamical behaviours --- the phases --- found in this work go beyond the LV-model itself and may play an important role in a variety of contexts from biology to economy,
which can be modeled by high-dimensional dynamical systems with random couplings.

\vspace{0.3cm}

\emph{Acknowledgments} - 
We acknowledge stimulating discussions with G. Bunin and G. Parisi on this subject.
This work was supported by the Simons Foundation Grant on \emph{Cracking the Glass Problem} (\# 454935 Giulio Biroli). 

\newpage

\appendix

\input{app-Ada}

\input{app-felix}

\bibliography{fullBibli_ada_felix}

\end{document}

%% file: app-Ada.tex
%\documentclass[10pt,a4paper]{article}
%\usepackage[utf8]{inputenc}
%\usepackage{amsmath}
%\usepackage{amsfonts}
%\usepackage{amssymb}
%\usepackage{graphicx}
%
%\usepackage{xcolor}		% For coloring math
%
%%\usepackage{natbib}
%\usepackage{bibentry}	% To cite full reference
%\nobibliography*
%
%\usepackage{caption}
%\usepackage{subcaption}
%
%\author{Felix Roy}
%
%\newcommand{\us}{$\_$}
%
%
%
%
%\begin{document}
%
%
%
%
%
%\title{Demographic noise in Lotka-Volterra\\Simulations for Ada}
%
%\author{Felix Roy, October 2019}
%
%\maketitle
%\tableofcontents
%

\newpage

\appendix

\section{Thermodynamics and replica formalism}

The evolution of the species abundances $N_i$ in the ecosystem (with $i=1,...,S$) is regulated by the following dynamical equation:
\begin{equation}
    \frac{d N_i}{dt}= -N_i \left[\nabla_{N_i} V_i(N_i) +\sum_{j, (j \neq i)} \alpha_{ij} N_j \right] +\sqrt{N_i} \eta_i(t) +\lambda
    \label{dynamical_eq}
\end{equation}
where $\eta_i(t)$ is a white noise with covariance $\langle \eta_i(t)\eta_j(t')\rangle=2 T\delta_{ij} \delta(t-t')$ and $T$ is the temperature. In the presence of a (demographic) noise -- an intrinsic population randomness due to birth, death and unpredictable interaction events -- Eq. (\ref{dynamical_eq}) represents a generalized Langevin equation with a one-species quadratic potential $V_i(N_i)$:
\begin{equation}
    V_i(N_i)= -\rho_i \left(K_i N_i- \frac{N_i^2}{2}\right) \ ,
\label{quadratic_potential}
\end{equation}
which allows us to precisely recover the well-known Lotka-Volterra equations \cite{Lotka1920, Volterra1927}.
The adimensional parameter $\rho_i=r_i/K_i$ denotes the ratio between the growth rate, $r_i$, and the carrying capacity, $K_i$.
For simplicity, we will assume no species dependence on these parameters and set $r_i=1$, $K_i=1$ in the following.

For a first-order analysis, the interaction matrix $\alpha_{ij}$ is assumed to be symmetric: its elements are i.i.d. Gaussian variables with mean and variance respectively:
\begin{equation}
\text{mean}[\alpha_{ij}]=\mu/S \hspace{0.5cm} \text{var}[\alpha_{ij}]=\sigma^2/S \ .  
\end{equation}
We can also suppose to add a small degree of asymmetry and perform a perturbative expansion. We should expect, even for arbitrary small asymmetry, that the critical (multiple equilibria) phase described in the main text will be suppressed, as proven long ago in spin-glass and neural network contexts \cite{Hertz1986, Hertz1987}.

In the case of symmetric interactions, the dynamical equation (\ref{dynamical_eq}) admits an invariant probability distribution in terms of an Hamiltonian operator $H$, as can be proven by writing the Fokker-Planck equation of the corresponding stochastic process. 
More precisely, we can safely define an Hamiltonian $H$ and solve the problem exactly within the replica formalism \cite{MPV}, as we will show in detail in the next Section.
We thus define
\begin{equation}
    H=\sum_{i} V_i(N_i) +\sum_{i<j}\alpha_{ij}N_i N_j +(T-\lambda) \sum_i \ln N_i
    \label{H_LV}
\end{equation}
where $\lambda$ denotes an infinitesimally small species-independent immigration rate. It essentially guarantees the existence of all invadable species. Furthermore, it has to satisfy the lower bound $\lambda$ with $\lambda>T$ such that the probability distribution $P(\lbrace N_i \rbrace)$ is correctly regularized at small $N_i$.

The connection with the statics is now clear: the original dynamical process in Eq. (\ref{dynamical_eq}) describes the time evolution of a large interacting ecosystem whose thermodynamics is determined by the Hamiltonian (\ref{H_LV}).
It is also worth noticing that in the limit $T \rightarrow 0$, $\lambda \rightarrow 0$, Eq.(\ref{H_LV}) reproduces precisely a spin-glass model \cite{MPV} where the continuous variables representing the abundances are then replaced by spin variables coupled by the $\alpha_{ij}$s. We will come back to this point in Appendix (\ref{SG-RR}).

\subsection{Replica symmetric ansatz}
\label{analysisRS}

To deal with the thermodynamics of disordered systems and to give a precise characterization of possible phase transitions associated with the emergence of complex collective behaviours we can resort to the replica method \cite{MPV, Parisi1983}. It was originally introduced to study spin-glass models but has now become a cornerstone for a vast class of complex systems.
The introduction of replicas allows us to handle quantities in which the disorder plays a key role and eventually obtain the free energy by means of the identity
\begin{equation}
    -\beta F=\lim_{n \rightarrow 0} \frac{\ln \overline{{Z^n}}}{n} \ .
\end{equation}
In principle, the free energy should depend on the specific realization of the disorder. However, in the thermodynamic limit we can safely ignore this issue since the free energy will converge to a unique value, thanks to its self-averaging property.
Operatively, one should compute the quantity on the r.h.s for integer values of the replica number $n$, then consider the analytical continuation to real values and only at the end of the computation take the limit $n \rightarrow 0$.

The starting point is then the computation of the replicated partition function, which in this specific case becomes:
\small{
\begin{equation}
    \overline{Z^n} = \overline{\int \prod_{i, (ij)} d N_i^a d \alpha_{ij} \exp \left[-\sum_{(ij)} \frac{(\alpha_{ij} - \mu /S)^2}{2 \sigma^2/S} -\beta H(\lbrace N_i^a \rbrace) \right]}
\end{equation}
where the overline denotes the average over the disorder, \emph{i.e.} the average over the Gaussian variables $\alpha_{ij}$. 
To perform the computation -- that will involve quadratic terms in the product of the species abundances -- we introduce the overlap matrix $Q_{ab}$ (with diagonal value $Q_{aa}$) and the external field $H_a$, where $(a,b)$ represent two replicas of the same system, \emph{i.e.}:
\begin{equation}
Q_{ab}=\frac{1}{S} \sum \limits_{i=1}^{S}  N_i^ a N_i^b \ ,
\end{equation}
\begin{equation}
    H_a=\frac{1}{S} \sum \limits_{i=1}^{S}  N_i^a  \ .
\end{equation}
We can thus rewrite the free energy in the replica space as:
\begin{equation}
 F= -\frac{1}{\beta n}\ln \overline{\int \prod \limits_{a,(a<b)} d Q_{ab}d Q_{aa} d H_{a} \; e^{S \mathcal{A}(Q_{ab}, Q_{aa},H_a) }}
 \label{freeenergy_RS}
\end{equation}
where the action reads
\begin{equation}
\begin{split}
& \mathcal{A}(Q_{ab},Q_{aa},H_a)=  -\rho^2 \sigma^2 \beta^2 \sum \limits_{a<b}\frac{Q_{ab}^2}{2}+\\
&-\rho^2 \sigma^2 \beta^2 \sum \limits_a \frac{Q_{aa}^2}{4}+\rho \mu \beta \sum_a \frac{H_a^2}{2}+\frac{1}{S}\sum \limits_i \ln Z_i \ .
\end{split}
\end{equation}
In turn, the partition function $Z_i$ is
\begin{equation}
Z_i= \int \prod \limits_a d N_i^a \exp \left(-\beta H_\text{eff}(\lbrace N^a \rbrace_i ) \right) \ ,
\end{equation}
which depends on the effective Hamiltonian:
\begin{equation}
\begin{split}
 H_{\text{eff}}(\lbrace N^a \rbrace_i )=& -\beta \rho^2 \sigma^2 \sum \limits_{a <b} N_i^a N_i^b Q_{ab} +\\
& -\beta \rho^2 \sigma^2 \sum \limits_a (N_i^a)^2 \frac{Q_{aa}}{2} + \sum \limits_a \rho \mu H_a N_i^a +\\
& + V_i(N_i^a) +(T-\lambda) \ln N_i^a \ .  
\end{split}
\end{equation}
The simplest scenario in the panorama of all possible replica techniques corresponds to the replica symmetric (RS) computation, which turns out to be correct as long as the free-energy landscape is characterized by one single equilibrium state. 
Any permutation of the replica indices does not affect the matrix structure. In order words, within the RS Ansatz the permutation symmetry of the replicated Hamiltonian is respected.

The overlap matrix is thus parametrized by two values: the self-overlap between replicas inside the same state, $q_d$, and the inter-state overlap, $q_0$. The external field is assumed to be uniform, $\forall a$.
\begin{equation}
\begin{split}
& Q_{ab}= q_0 \hspace{0.7cm }\text{if} \hspace{0.6cm} a \neq b    \\
& Q_{aa}=q_d \hspace{0.7cm }\text{if} \hspace{0.6cm} a = b\\
& H_a=h  \hspace{1.8cm } \forall a
\label{Ansatz_RS}
\end{split}
\end{equation}
The action $\mathcal{A}$ then becomes
\begin{equation}
\begin{split}
    \mathcal{A}(q_d,q_0,h)= & -\rho^2 \sigma^2 \beta^2 \frac{n(n-1)}{4} q_0^2 -\rho^2 \sigma^2 \beta^2 \frac{n}{4} q_d^2 +\\
    & + \rho \mu \beta \frac{n}{2} h^2 +\frac{1}{S} \sum \limits_{i} \ln Z_i 
    \end{split}
\end{equation}
where the partition function is integrated over $N_i^a$ with an effective Hamiltonian that depends now on the parameters $(q_d,q_0,h)$. 
Replica indices are nevertheless still coupled. 
At this stage, the replica trick comes into play allowing us to decouple replicas by the introduction of an auxiliary Gaussian variable $z$, with zero mean and unit variance, which makes the expression of the partition function of the form:
\begin{equation}
    Z_i= \int_{-\infty}^{+\infty} \frac{d z_i}{\sqrt{2 \pi}} e^{-z_i^2/2} \int \prod \limits_{a=1}^{n} d N_i^a e^{-\beta \sum \limits_a H_\text{RS}(N_i^a, z_i) } \ ,
\end{equation}
which is written in terms of the RS Hamiltonian:
\begin{equation}
\begin{split}
    H_\text{RS}(N_i,z_i)=&-\rho^2 \sigma^2 \beta (q_d-q_0) \frac{N_i^2}{2} +(\rho \mu h - z_i \rho \sqrt{q_0} \sigma)N_i+\\ & +V_i(N_i)+(T-\lambda)\ln N_i =\\
    & = \frac{N_i^2}{2}\left[\rho -\rho^2 \sigma^2 \beta(q_d-q_0)\right] +\\
    &+ \left(\rho \mu h -z_i \rho \sigma \sqrt{q_0} -\rho  \right)N_i +(T-\lambda)\ln N_i \ .
    \end{split}
    \label{H_RS}
\end{equation}
where we recall that $V_i(N_i)=-\rho N_i \left(1 - \frac{N_i}{2} \right)$ and the immigration rate $\lambda \rightarrow 0^{+}$.

In the thermodynamic limit, we can safely resort to the Laplace method and evaluate the integral by saddle-point approximation.
From the maximization of the action $\mathcal{A}(q_d,q_0,h)$, we get the corresponding for $(q_d,q_0,h)$, which, after considering the analytical continuation $n \rightarrow 0$, read:
\begin{equation}
\begin{split}
& q_d= \int \mathcal{D} z \left(\frac{ \int_{N_c}^{\infty} d N e^{-\beta H_{\text{RS}}(q_0,q_d,h,z)} N^2}{ \int_{N_c}^{\infty} d N e^{-\beta H_{\text{RS}}(q_0,q_d,h,z)}} \right) = \overline{\langle  N^2 \rangle} \ , \\ 
& q_0= \int \mathcal{D} z \left(\frac{ \int_{N_c}^{\infty} d N e^{-\beta H_{\text{RS}}(q_0,q_d,h,z)} N}{ \int_{N_c}^{\infty} d N e^{-\beta H_{\text{RS}}(q_0,q_d,h,z)}} \right)^2  = \overline{ \langle N \rangle^2}  \ , \\ 
& h= \int \mathcal{D} z \frac{ \int_{N_c}^{\infty} e^{-\beta H_{\text{RS}}(q_0,q_d,h,z)} N}{ \int_{N_c}^{\infty} d N e^{-\beta H_{\text{RS}}(q_0,q_d,h,z)}}=\overline{ \langle  N \rangle} \ .
\label{equations_log}
\end{split}
\end{equation}
where the calligraphic notation stands for the Gaussian integral $\mathcal{D}z \equiv \int \frac{dz}{\sqrt{2 \pi}} e^{-z^2/2}$.

More precisely, the most internal average corresponds to the standard average over the Boltzmann measure, while the external one corresponds to the average over the quenched disorder. 
A worth noticing aspect concerns the correct choice of the extremes for the integration over $N$. The integral cannot be extended over the interval $[0, \infty)$ and a more attentive analysis is needed because of the term $T \ln N$ in the Hamiltonian. The logarithmic term contributes to tilting the quadratic potential and providing a negative, divergent trend for values of the abundances very close to zero.
There are two alternatives: either play with the immigration rate $\lambda$ or carefully select the lower cut-off in the integral over the species abundances, $N_c$ (for more details, see Appendices \ref{Comparison-Num} and \ref{app:mathImmigration}).
We set the immigration rate to zero from the beginning of the computation. Conversely, we probe the optimal value of the cut-off $N_c$, eventually set to $10^{-2}$, as detailed also in \ref{matching}. Lower cut-off values cannot be selected as they would result in a non-optimal matching between our theoretical analysis and numerical simulations within a Dynamical Mean-Field Theory formalism.

The solution of Eqs. (\ref{equations_log}) is obtained by implementing an iterative algorithm. The iteration stops when the relative error between the value of the updated parameter and that at the previous time step is smaller than a given precision value, $\epsilon$. 
We study the convergence of the equations and their stability as a function of the demographic noise, whose amplitude corresponds to the physical temperature $T$.
The analysis at zero temperature has been recently performed in \cite{Biroli2018_eco}. Our work provides then a full and more comprehensive perspective. We derive the complete phase diagram starting from a high demographic noise initial condition -- corresponding to the high-temperature phase -- up to a very low demographic noise (corresponding to the zero-temperature limit).

We report below the resulting values at different inverse temperatures $\beta \equiv 1/T$ obtained via the numerical integration for $\mu=10$, $\sigma=1$ and a cut-off $N_c=10^{-2}$.
\begin{table}[h]
\centering
 \begin{tabular}{||c  c c c||} 
 \hline
 $\beta$ &  $q_d$ & $q_0$ & $h$ \\ [0.5ex] 
 \hline\hline
0.2 \hspace{0.35cm} & 0.391812 & 0.105042 & 0.323492 \\
1 \hspace{0.35cm} &  0.108159 & 0.0344899 & 0.185033  \\
5 \hspace{0.35cm} & 0.038878 &	0.014964 &	0.121134 \\
10 \hspace{0.35cm} &	  0.028545 & 	0.011976 &	0.107489  \\
15	\hspace{0.35cm} &     0.025080 & 	0.011114 &	0.102506  \\
20 \hspace{0.35cm}	&     0.023568 & 	0.010910 &	0.100248  	 \\
25	\hspace{0.35cm} &  	  0.022987 &	0.011080 &	0.099264	 \\
30	\hspace{0.35cm} &     0.023028 &	0.011568 &	0.099033 	\\
35 \hspace{0.35cm} &    0.023635 &	0.012435 &	0.099353 \\
40 \hspace{0.35cm} &    0.024938 &	0.013884 &	0.100173 \\
45 \hspace{0.35cm}  &  0.027021 & 0.016186 & 0.101385 \\
50 \hspace{0.35cm} &    0.031483 &	0.020585 &	0.103742 \\
55 \hspace{0.35cm} &    0.036909 &	0.026430 &	0.106083 \\
60 \hspace{0.35cm} &  0.042154 & 	0.032258 &	0.108096 \\
 \hline
 \end{tabular}
  \caption{Order parameters obtained by numerical integration for increasing values of the inverse temperature $\beta$ in the single equilibrium (RS) phase.}
  \label{table1}
\end{table}

Upon increasing $\beta$, the diagonal and off-diagonal value of the overlap matrix tend to become degenerate with $(q_d -q_0) \rightarrow 0$ (see Table \ref{table1}) and the stability matrix -- which is defined by the second derivative of the free energy with respect to the overlap -- develops zero modes, precisely at $\beta \approx 58$. The RS solution is then \emph{marginally stable}. In this regime, to correctly incorporate and describe the complexity of the landscape, a more structured Ansatz, so-called $1$RSB, must be introduced.

\subsection{One-step replica symmetry breaking ansatz}

In the presence of low demographic noise the RS solution is no longer appropriate to describe the thermodynamics of the system, which can be more correctly identified by a one-step replica symmetry breaking ($1$RSB) Ansatz.
This instability is intrinsically related to the emergence of multiple minima in the free-energy landscape, each of them associated with a different equilibrium configuration. 
According to the $1$RSB approximation, the $n$ replicas are now divided into $n/m$ different groups of $m$ replicas, with $0 \le m \le 1$ and $n/m$ integer. The overlap matrix assumes now three different values 
\begin{eqnarray}
Q_{ab} = 
  \begin{cases}
    q_d & \text{if} \hspace{0.4cm} a=b \\
    q_1 & \text{if} \hspace{0.4cm} a,b \in \mathcal{B}_l \\
    q_0 & \text{if} \hspace{0.4cm} a,b \not\in \mathcal{B}_l 
  \end{cases}
  \end{eqnarray}
  where $\mathcal{B}_l$ stands for the group of replicas in the same block.
This Ansatz yields the definition of an overlap matrix with $m-1$ off-diagonal elements equal to $q_1$ and $n-m$ elements equal to $q_0$.
A new parameter, $m$, comes into play which is usually denoted as \emph{breaking point parameter} in the replica jargon.
The external field is instead assumed to be uniform for all replicas $a$.
\begin{equation}
H_a=h \ , \hspace{1.cm} \forall a \ .
\end{equation}

Exactly as in Eq. (\ref{freeenergy_RS}), the free energy within the 1RSB Ansatz reads:
\begin{equation}
F^\text{$1$RSB}=-\frac{1}{\beta n} \ln \overline{ {\int dq_d d q_1 dq_0 d h \; e^{S \mathcal{A}(q_d,q_1,q_0,h)}}   }
\end{equation}
We are now able to write the resulting expression for the action $\mathcal{A}$:
\begin{equation}
\begin{split}
\mathcal{A}(q_d,q_1,q_0,h)=& -\rho^2 \sigma^2 \beta^2 \frac{n}{4}\left[(n-m)q_0^2+(m-1)q_1^2+q_d^2 \right]+\\
&+ \rho \mu \beta \frac{n}{2} h^2 +\frac{1}{S}\sum \limits_i \ln Z_i \ ,
\end{split}
\end{equation}
where the replicated partition function $Z_i$ turns out to be
\begin{equation}
\begin{split}
Z_i=& \int \frac{dz_i}{\sqrt{2 \pi}} e^{-z_i^2/2} \prod \limits_{a_B=1}^{n/m} \Biggl \lbrace \int \frac{dt_{a_B,i}}{\sqrt{2\pi}}  \prod \limits_{a \in a_B} d N_i^a \times \\
& \exp \left[-\frac{t_{a_B,i}^2}{2}-\beta \sum \limits_{a \in a_B} H_\text{$1$RSB}(N_i^a,z_i,t_{a_B,i})   \right] \Biggr \rbrace 
\end{split}
\end{equation}
hence, according to Eq. (\ref{H_RS}}), we have:
\begin{equation}
\begin{split}
    & H_\text{$1$RSB}(N_i,z_i,t_{a_B,i})= -\rho^2 \sigma^2 \beta (q_d-q_1) \frac{N_i^2}{2}+\\
    & + (\rho \mu h - t_{a_B,i}\rho \sigma \sqrt{q_1-q_0}-z_i \rho \sigma \sqrt{q_0} ) N_i +\\
    & + V_i(N_i)+ T \ln N_i \ .
    \end{split}
\end{equation}
To decouple replica indices according to a $1$RSB solution, we have thus introduced a double Gaussian integration leading to two different kinds of averages: i) over the single replica; ii) over a group of replicas in the same block of size $m$.
The two averages correspond respectively to
\begin{equation}
\langle \cdot \rangle_{\text{$1$r}}= \frac{\int_{N_c}^{\infty} d N \exp \left[-\beta H_{\text{$1$RSB}}(N,z,t_{a_B}) \right] \cdot}{\int_{N_c}^{\infty} d N \exp \left[-\beta H_{\text{$1$RSB}}(N,z,t_{a_B}) \right]}   \ ,
\label{average_1r}
\end{equation}
\begin{equation}
\langle \cdot \rangle_{\text{$m$-r}}= \frac{\int \frac{d t_{a_B}}{\sqrt{2 \pi}} e^{-\frac{t_{a_B}^2}{2}} \left(\int_{N_c}^{\infty} d N \exp \left[-\beta H_{\text{$1$RSB}}(N,z,t_{a_B}) \right] \right)^m \cdot}{\int \frac{d t_{a_B}}{\sqrt{2 \pi}} e^{-\frac{t_{a_B}^2}{2}} \left(\int_{N_c}^{\infty} d N \exp \left[-\beta H_{\text{$1$RSB}}(N,z,t_{a_B}) \right] \right)^m}  
\label{average_mr}
\end{equation}

The values of the overlap parameters $q_d$,$ q_1$, $q_0$ and the external field $h$ are obtained by saddle-point approximation of the action, leading to the following compact expressions
\begin{equation}
\begin{split}
& q_d= \overline{\langle \langle N^2 \rangle_\text{$1$r} \rangle_\text{$m$-r}} \\
& q_1=\overline{ \langle \langle N \rangle_\text{$1$r}^2 \rangle_\text{$m$-r}} \\
& q_0=\overline{ \langle \langle N \rangle_\text{$1$r} \rangle_\text{$m$-r}^2} \\
& h=\overline{ \langle \langle N \rangle_\text{$1$r} \rangle_\text{$m$-r}}
\label{optimal_RSB}
\end{split}
\end{equation}
that can be solved iteratively up to convergence. The initial condition is chosen close to the RS solution but with $q_1 > q_0$.
As the overlap parameters stand for the degree of similarity between two replicas in the same block or in different blocks, the found solution is thermodynamically relevant only if $q_d > q_1 > q_0$.
Indeed, in the very low-temperature limit, $q_d \rightarrow q_1$, similarly to the analysis we have performed for the RS solution (see \ref{analysisRS}).

The determination of the breaking parameter $m$ requires more attention. We can follow different routes depending on whether one aims to find the $m$-value that optimizes the expression of free energy or the one that identifies the aging solution in a dynamical framework. In this second case, the parameter $m$ is chosen in such a way that the found solution is marginal: one optimizes over the parameters in Eq. (\ref{optimal_RSB}) while selecting $m$ accordingly to the marginal stability condition.
We will devote the next Section to a detailed explanation of what \emph{marginal stability} means and what kind of consequences it yields for the system.

\subsection{Meaning of the replicon eigenvalue of the stability matrix}

To investigate the stability of the different phases, we introduce the Hessian matrix of the free energy, which allows us to study the harmonic fluctuations with respect to $\delta Q_{ab}$. Thanks to symmetry group properties of the replica space, the diagonalization of the stability matrix can be expressed in terms of three different sectors. Following \cite{Bray1979}, we define the three eigenvalues: the \emph{longitudinal}, $\lambda_\text{L}$, the \emph{anomalous}, $\lambda_\text{A}$, and the \emph{replicon}, $\lambda_\text{R}$.
We are specifically interested in the computation of the replicon mode as it is responsible for possible RSB effects. By contrast, a zero longitudinal mode can give information in terms of spinodal points describing how a state opens up along an unstable direction and originates then a saddle.

The detection of a vanishing replicon mode from the high-temperature (one single equilibrium) phase is related to the appearance of \emph{marginal states}. This feature is extremely important because of its intimate connection with out-of-equilibrium aging dynamics.

We consider then the variation of the RS action with respect to the overlap matrix
\begin{equation}
\begin{split}
 &\mathcal{A}(Q_{ab},Q_{aa},H_a)=  -\rho^2 \sigma^2 \beta^2 \sum \limits_{a<b}\frac{Q_{ab}^2}{2}-\rho^2 \sigma^2 \beta^2 \sum \limits_a \frac{Q_{aa}^2}{4}+\\
& +\rho \mu \beta \sum_a \frac{H_a^2}{2}+\frac{1}{S}\sum \limits_i \ln Z_i \ ,
\end{split}
\end{equation}
where the partition function
\begin{equation}
\begin{split}
Z_i& =  \int \prod_a d N_i^a  \exp \biggl [ \frac{\beta^2 \rho^2 \sigma^2}{2} \sum \limits_{a <b} Q_{ab} N_i^a N_i^b +\\
& \beta^2 \rho^2 \sigma^2 \sum \limits_a (N_i^a)^2 \frac{Q_{aa}}{2} -\rho \beta \mu \sum \limits_a N_i^a H^a -\beta V_i(N_i^a) -\ln {N_i^a} \biggr ]
 \end{split}
\end{equation}
which implies to the second order
\begin{equation}
\begin{split}
   \mathcal{M}_{abcd} & \equiv  -\frac{\partial^2 \mathcal{A}}{\partial Q_{ab}\partial Q_{cd}}=\\
   &=\beta^2 \rho^2 \sigma^2 \left[\delta_{(ab),(cd)}-(\beta^2 \rho^2 \sigma^2) \overline{\langle N^a N^b, N^c N^d \rangle_c}\right]
   \label{Mass}
   \end{split}
\end{equation}
where the subscript $\langle \cdot \rangle _c$ denotes the connected part of the correlator. 
When evaluated at the saddle point and in the limit $n \rightarrow 0$, the stability matrix (\ref{Mass}) can be decomposed as a function of three different correlators
\begin{equation}
\begin{split}
   \mathcal{M}_{abcd}=& M_{ab,ab} \left( \frac{\delta_{ac}\delta_{bc}+\delta_{ad}\delta_{bc}}{2}\right) +\\
   +& M_{ab,ac} \left( \frac{\delta_{ac}+\delta_{bd}+\delta_{ad} +\delta_{bc}}{4} \right)+\\
   + & M_{ab,cd}
   \end{split}
\end{equation}
from which the projection on the replicon subspace is
\begin{equation}
 \lambda_\text{R}=   (\beta \rho \sigma)^2  \biggl [ 1-(\beta \rho \sigma)^2 \overline{\left( M_{ab,ab}-2M_{ab,ac}+M_{ab,cd} \right) } \biggr ]
\end{equation}
The $\langle \cdot \rangle$ is performed over the effective Hamiltonian, while the average over the quenched disorder is always denoted as $\overline \cdot$. The replicon mode gives information about the fluctuations inside the innermost block of the overlap matrix, corresponding to the fluctuations within the same state. According to this structure and the different combinations of the replica indices, three elements must be determined
\begin{equation}
\begin{split}
&  M_{ab,ab}-2M_{ab,ac}+M_{ab,cd} =\\
&    \left[ \langle (N^a)^2(N^b)^2\rangle -2 \langle (N^a)^2 N^b N^c \rangle +\langle N^a N^b N^c N^d \rangle \right]
    \end{split}
\end{equation}

\begin{figure}[h]
\includegraphics[scale=0.33]{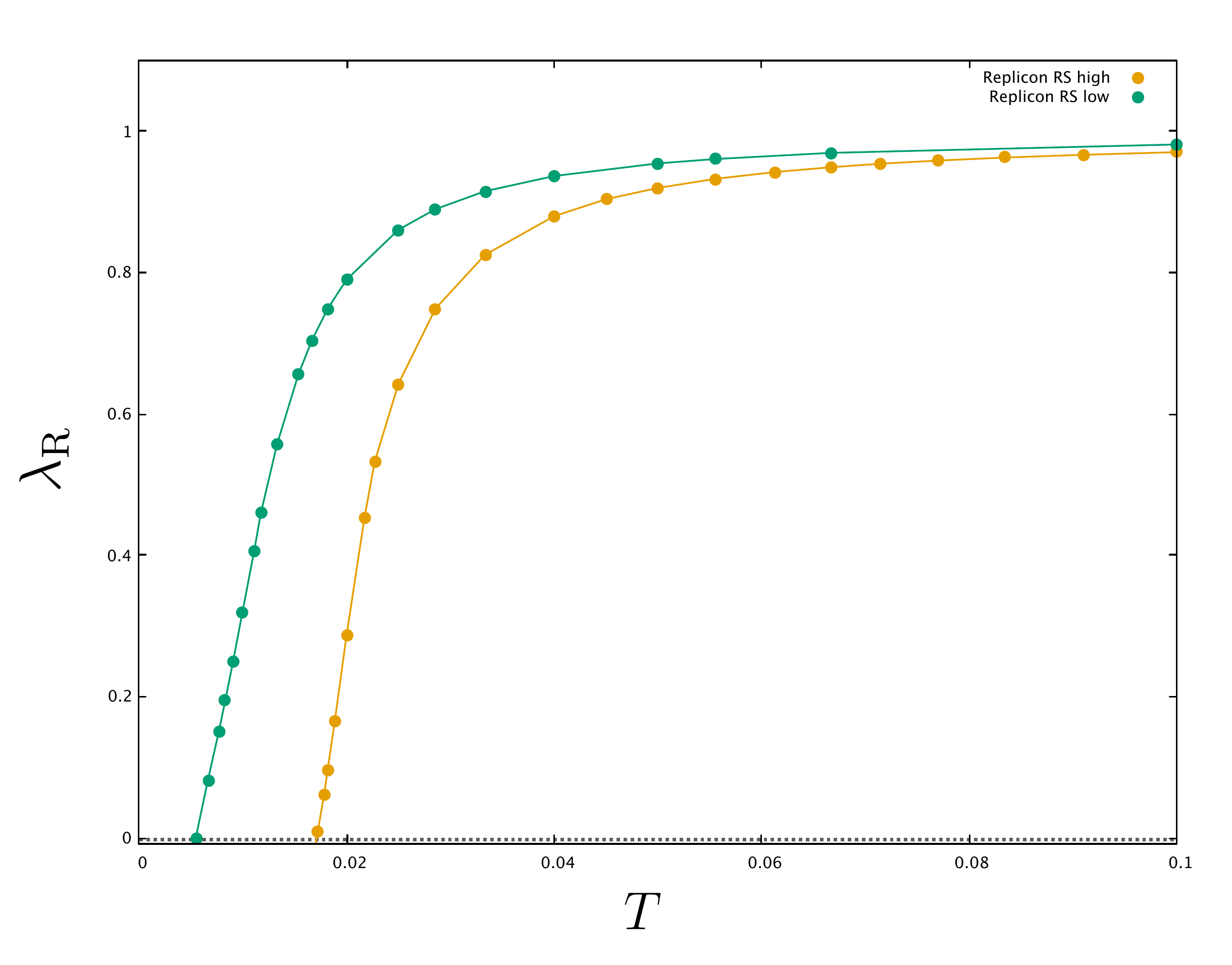}
\caption{Replicon eigenvalue obtained within the replica symmetric (RS) Ansatz for $\mu=10$, $\sigma=1$ (in orange) and $\sigma=0.85$ (in green) as a function of temperature. Below the dashed black line corresponding to the zero value, the replicon becomes negative and makes the RS approximation no longer valid. Upon decreasing $\sigma$, the transition towards the $1$RSB phase occurs at a lower critical temperature.}
\end{figure}

In the simplest scenario corresponding to the presence of one single equilibrium, the expression for the replicon eigenvalue can be further simplified as:
\begin{equation}
\lambda_\text{R}=(\beta \rho \sigma)^2 \left[ 1-(\beta \rho \sigma)^2 \overline{ \left(\langle N^2 \rangle - \langle N \rangle^2 \right)^2} \right] \ ,
\end{equation}
where the averaged difference describes the fluctuations between the first and second moment of the species abundances within one state, namely between the diagonal value $q_d$ and the off-diagonal contribution $q_0$ of the overlap matrix. The difference between the two overlap values can also be interpreted as the response function of the single species to an infinitesimal perturbation.
A large, diverging response -- which is related to the inverse of the replicon mode -- is a further evidence that the system is close to a critical point. 

We can repeat exactly the same procedure to obtain the replicon in the $1$RSB Ansatz. The analysis of harmonic fluctuations proceeds exactly as for the RS case. The only difference now is that -- as for Eqs. (\ref{average_1r})-(\ref{average_mr}) -- we have to distinguish between the intra-state average and the inter-state average, that is over a bunch of replicas in the inner block of size $m$.
\begin{equation}
\lambda_\text{R}^\text{1rsb}=(\beta \rho \sigma)^2 \left[ 1-(\beta \rho \sigma)^2 \overline{ \langle \left(\langle N^2 \rangle_\text{$1$r} - \langle N \rangle^2_\text{$1$r} \right)^2 \rangle_\text{$m$-r}} \right] \ ,
\end{equation}
The value at which the $1$RSB replicon vanishes signals a critical phase transition towards a more structured phase. This so-called \emph{Gardner phase} turns out to be characterized by a hierarchical organization of the different equilibria, whose analytical solution is well described within a Full RSB Ansatz. For the first time in an ecological context we are able to prove in a rigorous way the analogy between glassy physics and complex energy landscapes in large ecosystems, with respect to previous studies in this direction \cite{Biroli2018_eco, Altieri2019}.

\section{Derivation of the complete phase diagram}

Our analysis, based on increasingly structured Ansatz, has allowed us to derive the complete phase diagram of the Lotka-Volterra model as a function of the demographic noise. 
By varying this control parameter, we have highlighted the existence of different phase transitions of increasing complexity.
Thus far the only available results have been obtained without demographic noise, exactly at zero temperature \cite{Bunin2017}.

\begin{figure}[h]
\includegraphics[scale=0.32]{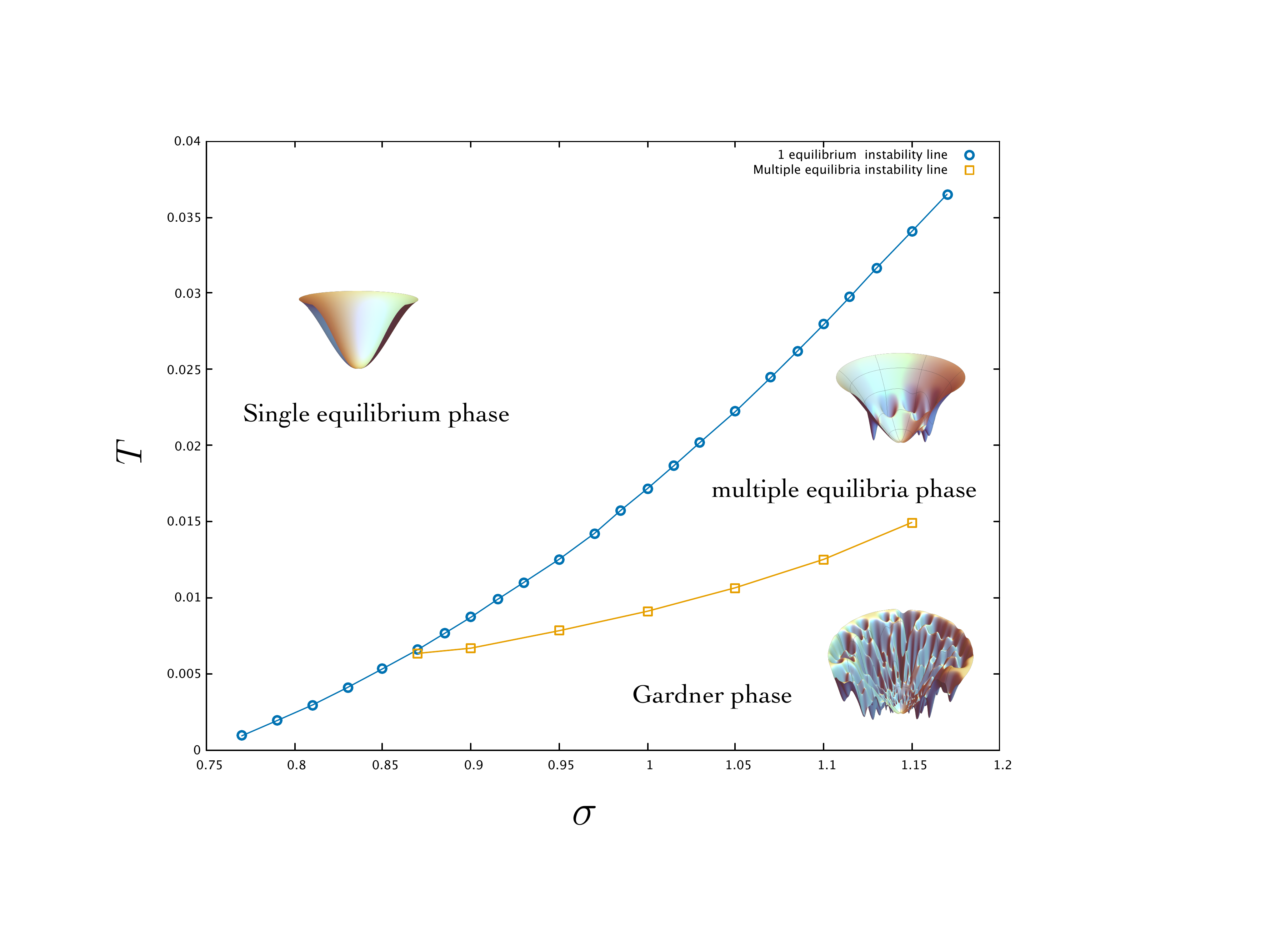}
\caption{Two-dimensional phase diagram of the Lotka-Volterra model in the presence of demographic noise as a function of the heterogeneity parameter $\sigma$ (in the presence of the logarithmic term and with a selected value of the cutoff $N_c=10^{-2}$). We can distinguish three different phases: i) a single equilibrium phase above the blue line; ii) a multiple equilibria regime ($1$RSB stable phase) between the light blue and the orange lines; iii) a Gardner phase, characterized by a hierarchical organization of the different equilibria in the free-energy landscape.}
\label{Phase_diag_complete}
\end{figure}

In correspondence of the blue line in Fig. (\ref{Phase_diag_complete}) the landscape structure is no longer identified by a single equilibrium but should be replaced by a two-level hierarchy, which results in the appearance of new equilibrium configurations.

In particular we have realized that with very low demographic noise the system undergoes a Gardner transition to a new marginally stable phase: each amorphous state, say a basin, is fragmented into a fractal structure of sub-basins (\emph{metabasin}). The internal structure of states in which a basin splits is described by the Full RSB solution of the partition function, according to which the replica symmetry is broken an infinite number of times. Similar to what observed in low-temperature glasses, such a transition will revolutionize our understanding of ecosystems.
The emergence of a infinite hierarchy of scales for the order parameter is also related to an infinite number of time-scales playing a central role in population dynamics, as we will discuss in the next Sections.

What a Full RSB regime precisely yields for biological and ecological settings with the appearance of an abundance of soft modes is particularly fascinating and still an open matter of debate.

\section{Results for the spin-glass model without logarithmic interaction}
\label{SG-RR}

We consider exactly the same system without the logarithmic interaction in the species abundances. In the limit $\lambda \rightarrow 0^{+}$, the resulting Hamiltonian is
\begin{equation}
    H=\sum_{i} V_i(N_i) +\sum_{ij}\alpha_{ij}N_i N_j
    \label{H_spinglass}
\end{equation}
where the first term has the same quadratic dependence in the species abundances as in Eq. (\ref{quadratic_potential}), while the latter represents the pairwise interacting part of the Hamiltonian.
In theoretical physics, Eq. (\ref{H_spinglass}) is an example of \emph{spin-glass} model where the degrees of freedom can be either discrete variables, to model magnetic spins, or continuous variables. The couplings between spins are chosen randomly, taking both positive and negative values and thus resulting in frustration and non-convex optimization phenomena.
The simplest and most classical example of spin glass is represented by the Sherrington-Kirkpatrick (SK) model \cite{SK1975} where each spin variable interacts with all the others in a fully-connected topology. 

\subsection{Connections with the random replicant model}

In theoretical ecology, a similar model was first introduced by Diederich and Opper in the 80s, studying the mean-field dynamical evolution of $S$ randomly interacting species with a deterministic self-interaction \cite{Diederich1989replicators}. This pioneering model is usually known as \emph{replicator model}. Later on, a detailed study was also performed in \cite{Biscari1995} by using the replica formalism in the zero-temperature limit.
Even though without the demographic noise, this model turns out to be not particularly interesting in a purely ecological context, it can still provide further insights on the replicator equations in many other interdisciplinary domains \cite{Cressman2014}, such as in the study of Nash equilibria \cite{Berg1999} in Game Theory.

Given $S$ species, the Lyapunov function of the replicator model is written as a function of the concentration variables $x_i$ in the following way
\begin{equation}
\mathcal{H}_{\text{R}}= \sum \limits_{i< j=1}^{S} J_{ij}x_i x_j -a \sum \limits_{i=1}^{S}x_i^2
\end{equation}
where the control parameter $a$ limits the growth rate of one single species. The global constraint contributes to non-convex constraint optimization properties with non-trivial symmetry breaking effects.

The couplings $J_{ij}$ are extracted from a Gaussian distribution with zero mean and variance $J^2$, \emph{i.e.}:
\begin{equation}
    P(J_{ij})=\sqrt{\frac{S}{\pi J^2}} \exp \left( -\frac{S J_{ij}^2}{J^2} \right)
\end{equation}
playing the same role as the $\alpha_{ij}$s in our notation. $x_i$ are real variables subject to the global constraint $\sum_i x_i=S$,  $\forall x_i \ge 0$ with $i=1,...,S$.
One can introduce a Lagrange multiplier $\gamma$ to enforce the normalization condition over $x_i$, which leads to the following expression for the partition function:
\begin{equation}
\begin{split}
Z=\int_{0}^{S}\prod_{i=1}^{S}   d x_i \; \int_{-i\infty}^{i \infty} d \gamma \exp \biggl [ & -\beta \sum_{ij} J_{ij}x_ix_j -\beta a\sum_i x_i^2 +\\
& -\gamma \sum_i (x_i-1) \biggr ] \ .
\label{Z_nRRM}
\end{split}
\end{equation}
As usual, one can introduce the replica trick to average over the quenched variables $J_{ij}$ and thus rewrite Eq. (\ref{Z_nRRM}) as
\begin{equation}
\begin{split}
    Z_n=\int d\gamma_\alpha \text{Tr}_{n S} \exp \biggl [ & -S \sum_{\alpha} \left(\beta a X_{\alpha}^2+\gamma_{\alpha}(X_{\alpha}-1)\right) +\\
    & +\frac{S^2 \beta^2 J^2}{4} \sum_{\alpha \beta} X_{\alpha}^2X_{\beta}^2
    \biggr]
    \end{split}
\end{equation}
where $X$ are exactly the same $n$ variables as before that have now lost their dependence on the index $i$ because of the equivalence of all sites. By introducing the overlap parameter $Q_{\alpha \beta}$ between $X_{\alpha}$ and $X_{\beta}$, one obtains
\begin{equation}
\begin{split}
Z_n=\int d \gamma_\alpha d Q_{\alpha \beta} & \exp \left(-S \sum_{\alpha \beta} Q_{\alpha \beta}^2+S\sum_\alpha \gamma_\alpha \right)\times\\
& \times \text{Tr}_{n S} \exp{ \left[ S \mathcal{L}(Q,\gamma, X) \right] }
\end{split}
\end{equation}
where the action $\mathcal{L}(Q,\gamma,X)$ is
\begin{equation}
\mathcal{L}(Q,\gamma, X)\equiv \beta J \sum_{\alpha \beta} Q_{\alpha \beta}X_\alpha X_\beta  -\beta a \sum_\alpha X_\alpha^2-\sum_\alpha \gamma_\alpha X_\alpha \ .
\end{equation}
The second term, absent in the ordinary SK model, is actually due to the global quadratic constraint on the species concentrations. 
Using the same Ansatz as in \cite{Biscari1995} and optimizing over $Q_{\alpha \beta}$ and $\gamma_\alpha$ with the following choice:
\begin{equation}
\begin{cases}
& Q_{\alpha \beta}=q \delta_{\alpha \beta} +t\\
& \gamma_\alpha=\gamma
\end{cases}
\end{equation}
one can get to writing the replica symmetric action 
\begin{equation}
\mathcal{L}_{\text{RS}}(q,t,\gamma,X)=-\beta (a-qJ) \sum_\alpha X_\alpha^2+\left( 2 z \sqrt{\beta J t}-\gamma \right)\sum_\alpha X_\alpha  
\end{equation}
where the auxiliary Gaussian variable $z$ has been introduced to decouple replicas.
Because under this transformation the integrals become independent, one can safely forget the dependence on the replica index $\alpha$ and re-scale the expression above by $J$
\begin{equation}
\mathcal{L}_{\text{RS}}(q,t,\gamma,X)=-\beta J\left[ \left( a/J-q\right)X^2-\left(2 z \sqrt{\tilde{t}}-\tilde{\gamma} \right)X \right]
\end{equation}

According to our notation and in particular to Eqs. (\ref{Ansatz_RS})-(\ref{H_RS}) considered without the logarithmic term and the immigration parameter, a direct mapping between the two models can be immediately pointed out. We additionally consider a quadratic potential quadratic potential $V(X)=-\rho X \left(1 - \frac{X}{2} \right)$ with respect to the original model proposed in \cite{Biscari1995}, where we have replaced $N \leftrightarrow X$ for consistency of notation. We can thus conclude 
\begin{equation}
\begin{split}
& \left[\rho -\rho^2 \sigma^2 \beta(q_d-q_0)\right] \Longleftrightarrow -\beta J (a/J-q)\\
    & \left(\rho \mu h -z \rho \sigma \sqrt{q_0} -\rho  \right) \Longleftrightarrow  -2 z\sqrt{\tilde{t}}-\tilde{\gamma}  \ .
    \label{mapping_RRM}
    \end{split}
\end{equation}
The Lotka-Volterra model embeds a quadratic one species potential as a function of $N_i^2$, which contributes to shifting the linear and quadratic terms of an amount $\rho$. The parameter $\tilde{\gamma}$ in the random replicant model is related to the (competitive) interaction term $\mu$, while $\tilde{a}\equiv a/J$ is associated with the heterogeneity parameter $\sigma$.

In \cite{Biscari1995} it was shown there exists a critical value of the control parameter $\tilde{a}=\equiv \frac{a}{J}$, $\tilde{a}_c=1/\sqrt{2}$, below which the system develops non-ergodic features and replica symmetry breaking effects. 
At higher values, irrespective of the initial conditions, only one equilibrium is possible, whereas in the region characterized by $\tilde{a}<\tilde{a}_c$ the final configuration is strongly affected by a small perturbation that can allow the system to reach different equilibria. This last phase corresponds to a high-competition scenario between the interacting species.
Interestingly, the replicator model can be proven to have a one-to-one mapping with many different problems of interest in optimization and disordered systems. 

\subsection{Exact solution in the replica symmetric case}

The following analysis focuses on the Lotka-Volterra Hamiltonian, as reported in Eq. (\ref{H_LV}), and considered in the absence of the logarithmic term. We will derive exact expressions for the order parameters to be analyzed as a function of temperature up to $T=0$. In this case, the temperature explicitly comes out in the weighting factor $\exp(-\beta H)$ and implies for the replicated partition function:
\begin{equation}
    \overline{Z^n} = \overline{\int \prod_{i, (ij)} d N_i^a d \alpha_{ij} \exp \left(-\sum_{(ij)} \frac{(\alpha_{ij} - \mu/ S)^2}{2 \sigma^2/S} -\beta H(\lbrace N_i \rbrace) \right)}
\end{equation}

Exactly as for the previous analysis, we have performed both a RS and $1$RSB computation.
In the former, the saddle-point equations read:
\begin{equation}
\begin{split}
& q_d= \int \mathcal{D} z \left(\frac{ \int_{0}^{\infty} d N e^{-\beta H_{\text{RS}}(q_0,q_d,h,z)} N^2}{ \int_{0}^{\infty} d N e^{-\beta H_{\text{RS}}(q_0,q_d,h,z)}} \right) \ , \\ 
& q_0= \int \mathcal{D} z \left(\frac{ \int_{0}^{\infty} d N e^{-\beta H_{\text{RS}}(q_0,q_d,h,z)} N}{ \int_{0}^{\infty} d N e^{-\beta H_{\text{RS}}(q_0,q_d,h,z)}} \right)^2  \ , \\ 
& h= \int \mathcal{D} z \frac{ \int_{0}^{\infty} e^{-\beta H_{\text{RS}}(q_0,q_d,h,z)} N}{ \int_{0}^{\infty} d N e^{-\beta H_{\text{RS}}(q_0,q_d,h,z)}} \ .
\label{no-cutoff}
\end{split}
\end{equation}
\vspace{1cm}

where now the integral over $N$ is extended over the entire positive axis without fixing any cut-off. $\mathcal{D}z$ denotes, as usual, the Gaussian integration over the auxiliary variable $z$ -- which is introduced to decouple replicas -- with zero mean and unit variance. 
The RS Hamiltonian then reads:
\begin{equation}
\begin{split}
    H_\text{RS}(N,z)=&-\rho^2 \sigma^2 \beta (q_d-q_0) \frac{N^2}{2} +(\rho \mu h - z \rho \sqrt{q_0} \sigma)N+ V(N) \\
    =& \frac{N}{2}\left[\rho -\rho^2 \sigma^2 \beta(q_d-q_0)\right] + \left(\rho \mu h -z \rho \sigma \sqrt{q_0} -\rho K  \right)N  \ .
    \end{split}
\end{equation}
Therefore, Eqs. (\ref{no-cutoff}) can be exactly computed in terms of the error function and its combinations, {\it i.e.}
\begin{widetext}
\begin{equation}
    \begin{split}
 q_d=\int \mathcal{D}z & \Biggl \lbrace \Biggl [ e^{-\frac{\beta \rho (K-h\mu+\sqrt{q_0} z \sigma)^2}{2-2(q_d-q_0)\beta \rho \sigma^2}} \Biggl ( 2(K -h\mu +\sqrt{q_0} z \sigma) \sqrt{\beta \rho[1-(q_d-q_0)\beta \rho \sigma^2]}+ \\
& e^{\frac{\beta \rho (K-h\mu+\sqrt{q_0} z \sigma)^2}{2-2(q_d-q_0)\beta \rho \sigma^2}} \sqrt{2 \pi} \biggl [ 1+\beta \rho \left[ (K-h\mu)^2 +2 \sqrt{q_0} z(K-h\mu)\sigma +(q_0-q_d+q_0z^2) \sigma^2 \right]  \biggr ] \times \\
& \times \biggl [ 1+ \text{Erf}\left( \frac{\beta \rho (K - h \mu + \sqrt{q_0} z \sigma)}{\sqrt{2} \sqrt{\beta \rho(1 + q_0 \beta \rho \sigma^2 -q_d \beta \rho \sigma^2)}}  \right) \biggr ] \Biggr ) \Biggr ]\frac{1}{\sqrt{2} Z} \Biggr \rbrace \ ,
\label{qd-nocutoff}
\end{split}
\end{equation}
\\
\begin{equation}
\begin{split}
 q_0= \int \mathcal{D}z & \Biggl \lbrace \Biggl [  e^{-\frac{\beta \rho (K-h\mu+\sqrt{q_0} z \sigma)^2}{2-2(q_d-q_0)\beta \rho \sigma^2}} \Biggl ( 
e^{\frac{\beta \rho (K-h\mu+\sqrt{q_0} z \sigma)^2}{2-2(q_d-q_0)\beta \rho \sigma^2}} \sqrt{\pi} \beta \rho (K-h\mu +\sqrt{q_0} z \sigma) + \sqrt{2 \beta \rho (1+ q_0\beta \rho\sigma^2-q_d \beta \rho \sigma^2)}+ \\
&+ e^{\frac{\beta \rho (K-h\mu+\sqrt{q_0} z \sigma)^2}{2-2(q_d-q_0)\beta \rho \sigma^2}}  \sqrt{\pi} \beta \rho \left( K-h\mu +\sqrt{q_0} z \sigma \right) \text{Erf}\left[\frac{\beta \rho (K - h \mu + \sqrt{q_0} z \sigma)}{\sqrt{2} \sqrt{\beta \rho(1 + q_0 \beta \rho \sigma^2 -q_d \beta \rho \sigma^2)}}  \right]  \Biggr ) \Biggr ]  \frac{1}{Z} \Biggr \rbrace^2 \ ,  
\label{q0-nocutoff}
\end{split}
\end{equation}
\\

\begin{equation}
    \begin{split}
         h=\int \mathcal{D}z  & \biggl [ \sqrt{\pi} \beta \rho(K-h\mu +\sqrt{q_0} z \sigma) + \sqrt{2} e^{-\frac{\beta \rho (K-h\mu+\sqrt{q_0} z \sigma)^2}{2-2(q_d-q_0)\beta \rho \sigma^2}} \sqrt{\beta \rho (1+ q_0\beta \rho \sigma^2-q_d \beta \rho \sigma^2)}+\\
        & + \sqrt{\pi} \beta \rho (K-h\mu +\sqrt{q_0} z \sigma) \text{Erf}\biggl [ \frac{\beta \rho (K - h \mu + \sqrt{q_0} z \sigma)}{\sqrt{2} \sqrt{\beta \rho(1 + q_0 \beta \rho \sigma^2 -q_d \beta \rho \sigma^2)}}  \biggr ] \biggr]  \frac{1}{Z} \ ,
        \label{h-nocutoff}
    \end{split}
\end{equation}

where the partition function reads:
\begin{equation}
Z= \sqrt{\pi} \beta \rho [1-(q_d-q_0) \beta \rho \sigma^2]^2 \biggl [ 1+ \text{Erf}\left( \frac{\beta \rho (K - h \mu + \sqrt{q_0} z \sigma)}{\sqrt{2} \sqrt{\beta \rho(1 + q_0 \beta \rho \sigma^2 -q_d \beta \rho \sigma^2)}}  \right) \biggr ] \ .
\end{equation}
\end{widetext}
Exactly as for the previous Section where we have derived the expression for the replicon eigenvalue, we have here:
\begin{equation}
\lambda_\text{R}=(\beta \rho \sigma)^2 \left[ 1-(\beta \rho \sigma)^2 \overline{ \left(\langle N^2 \rangle - \langle N \rangle^2 \right)^2} \right] \ ,
\end{equation}
where the expressions for the averaged values $\langle N^2\rangle$ and $\langle N \rangle^2$ correspond precisely to $q_d$ and $q_0$ obtained in Eqs. (\ref{qd-nocutoff})-(\ref{q0-nocutoff}) according to a RS Ansatz. The overline denotes the average over the quenched disorder, which is technically implemented by integrating over the Gaussian variable $z$.

\subsection{Replica symmetry broken case}

The computation can be easily generalized to more complex scenarios. In particular, in the $1$RSB case the Hamiltonian reads 
\begin{equation}
\begin{split}
    H_\text{$1$RSB}(N,z,t)=&  \frac{N^2}{2}
\left[\rho -\rho^2\sigma^2 \beta (q_d-q_1) \right]+\\
+& (\rho \mu h - t \rho \sigma \sqrt{q_1-q_0} -z \rho \sigma \sqrt{q_0} -\rho K)N \\
    \end{split}
\end{equation}
with the only difference of considering now the double Gaussian integration over $z$ and $t$ that account respectively for the single replica average and the $m$-block average. Even though the structure of the resulting equations for $(q_d,q_1,q_0,h)$, we are still able to get an exact closed-form derivation. Exactly as before for the most general case, the saddle-point equations read 
\begin{equation}
\begin{split}
& q_d= \overline{\langle \langle N^2 \rangle_\text{$1$r} \rangle_\text{$m$-r}} \\
& q_1=\overline{ \langle \langle N \rangle_\text{$1$r}^2 \rangle_\text{$m$-r}} \\
& q_0=\overline{ \langle \langle N \rangle_\text{$1$r} \rangle_\text{$m$-r}^2} \\
& h=\overline{ \langle \langle N \rangle_\text{$1$r} \rangle_\text{$m$-r}}
\label{optimal_1RSB}
\end{split}
\end{equation}
to be solved iteratively.

\subsection{Derivation of the phase diagram: evidence of a Gardner phase}

By fixing the carrying capacity $K=1$ and $\rho=1$ in Eqs. (\ref{qd-nocutoff})-(\ref{q0-nocutoff})-({\ref{h-nocutoff}), we have explored the phase space as a function of the parameter $\beta=1/T$ along with the variation of the mean and the variance $\mu$ and $\sigma$ respectively of the interaction matrix. The equations (\ref{no-cutoff}) have been solved iteratively starting from the high-temperature region up to zero temperature. For high values of $\sigma$ and $T$ there exists only one equilibrium: the system is thus able to recover its equilibrium configuration even starting from different initial conditions. As shown in Fig. (\ref{PD_nolog}), upon decreasing $T$ new phases emerge, leading respectively to an intermediate stable $1$RSB phase and a Gardner phase in the very low-temperature limit. 

We have thus obtained the complete phase diagram at fixed $\mu=10$.

\begin{figure}[h]
    \centering
    \includegraphics[scale=0.29]{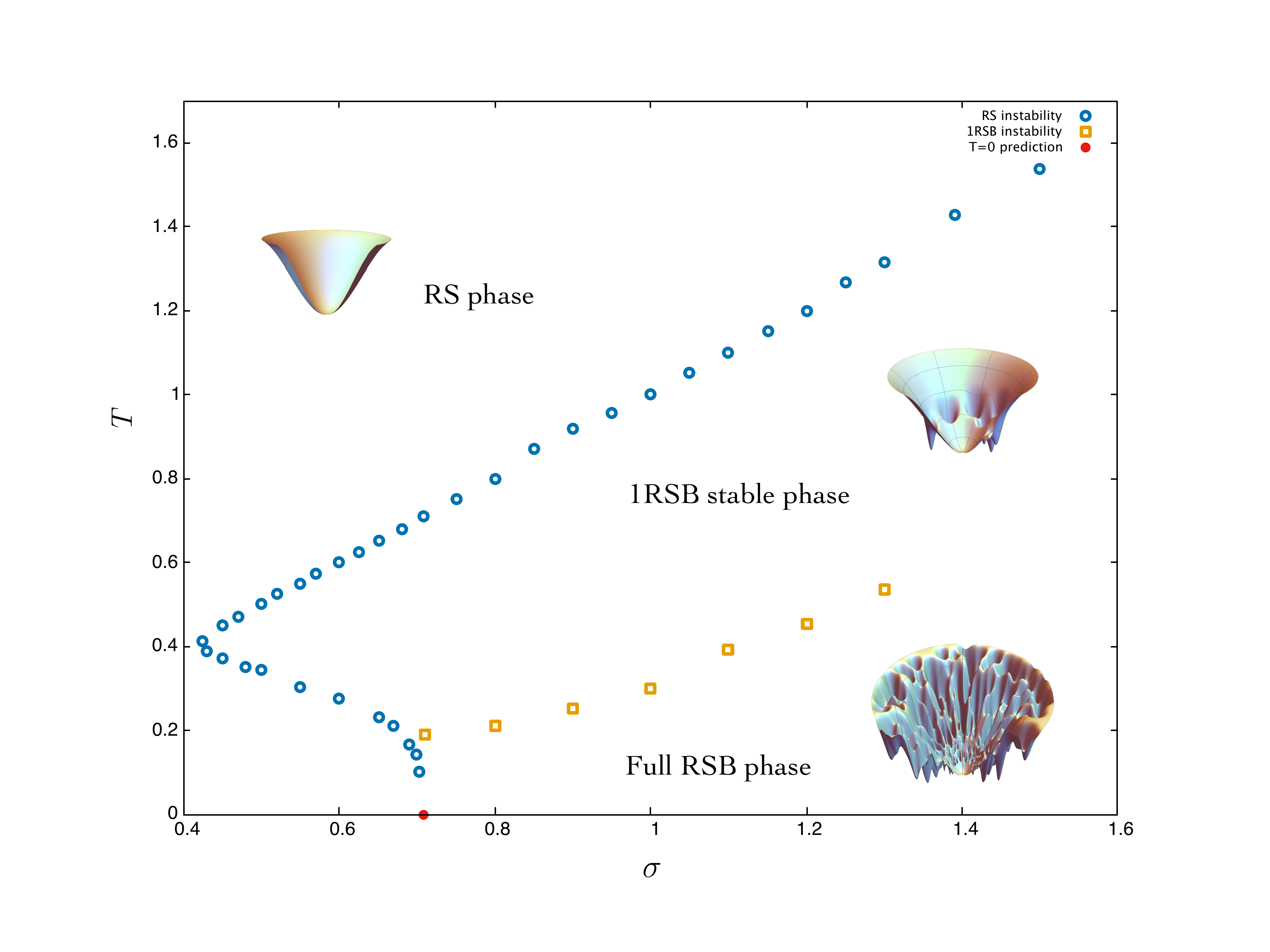}
    \caption{Two-dimensional phase diagram showing the instability lines respectively for the one equilibrium (RS) phase and the multiple equilibria phase ($1$RSB) as a function of $\sigma$ (without the logarithmic term). Between the two lines (in light blue and orange respectively) a stable $1$RSB phase persists. The red dot corresponds to the analytically predicted value in the zero-temperature limit allowing for the estimation of the critical value $\sigma_c=1/\sqrt{2} \approx 0.707$.}
    \label{PD_nolog}
\end{figure}
In Eqs. (\ref{mapping_RRM}) we have derived the exact mapping between our model and the Random Replicant Model (RRM), pointing out how the mean and the variance of the interaction matrix are essentially related to the Lagrange multiplier $\tilde{\gamma}$ and the parameter $\tilde{a}$ ensuring the global quadratic constraint. 
In our case the phase diagram does not display any special dependence on the mean interaction $\mu$ and can be thus analyzed at fixed value, by changing only the heterogeneity and the demographic noise. 
However, at variance with the original RRM model, fixing $\mu$ would correspond to allowing the sum of the species concentrations to vary, \emph{i.e.} injecting or removing species in the ecosystem.

\section{Derivation of the complexity}

We aim to determine the number of local minima $\mathcal{N}$ of the free energy corresponding to the different equilibrium configurations in the landscape structure. The logarithm of the number of minima divided by $S$ defines the complexity $\Sigma(f)$ (a.k.a. \emph{configurational entropy}) of the ecosystem. This provides a crucial information to determine what kind of universality class the system belongs to. In particular, in statistical physics of disordered systems two main universality classes are well-known: i) spin-glass models, in which the number of minima of the free energy is not exponential in the system size and the free energy barriers are sub-extensive; ii) structural glasses, for which the number of free-energy minima is actually exponential in the system size. Then, there exists a finite temperature range, between a static transition temperature $T_s$ and a dynamical transition temperature $T_d$, within which the complexity turns out to be finite. 

As we deal with mean-field models, we can resort to a simplified description for which the total partition function of the model can be expressed as the sum of $\alpha$ \emph{pure states} contributions. More precisely, we can define a generic effective potential as a function of local order parameters whose minima are in correspondence one-to-one with the so-called Thouless-Anderson-Palmer states \cite{TAP, MPV, Crisanti1995_TAP, Cavagna1998}. 
Provided the number of minima of the free energy is exponential in the system size, the partition function can be expressed as:
\begin{equation}
\begin{split}
Z \sim & \sum \limits_{\alpha} e^{-\beta N f_{\alpha}(T)}=\int d f \; \sum \limits_\alpha \delta(f-f_\alpha(T))e^{-\beta N f}= \\
& \int d f \rho(f) e^{-\beta N f} \sim e^{N \left[ \Sigma(f^{*},T)-\beta f^{*}\right]}n \ .
\end{split}
\end{equation}
Therefore, the configurational entropy corresponds to
\begin{equation}
    \Sigma(f,T) \equiv \frac{1}{S} \ln \sum \limits_{\alpha} \delta(f-f_\alpha(T))
\end{equation}
while $f^{*} \in \left[ f_0,f_\text{th}\right]$ satisfies the extremal condition:
\begin{equation}
\frac{d}{d f} \left[f-(1/\beta) \Sigma(f,T) \right]=0
\end{equation}
namely
\begin{equation}
\frac{d \Sigma(f,T)}{d f}=\frac{1}{T}\ .
\label{complexity_min}
\end{equation}
Three different regimes can be generically observed in mean-field models:
\begin{itemize}
    \item i) in the high-temperature phase (corresponding to high demographic noise), namely for above the dynamical transition temperature, the paramagnetic state dominates the free-energy density for any value $\left[f_0,f_\text{th} \right]$;
    \item ii) between the dynamical and the statical transitions, there exists a value $f^{*}$ such that the quantity $f^{*}-(1/\beta) \Sigma(f^{*})$ evaluated in $f^{*}$ coincides with the paramagnetic value of the free-energy density. However, in this second regime, the resulting state is composed by an exponential number of metastables states of individual free-energy density $f^{*}$.
    Upon crossing the dynamical temperature, the free energy preserves its analyticity without undergoing any true phase transition;
    \item iii) For temperatures lower the static transition temperature, the leading contribution is due to the lowest free-energy states with $f^{*}=f_0$ and, as a consequence, $\Sigma(f_0)=0$. In this phase, the number of states is sub-exponential in the system size.
    \end{itemize}
While in the very high-temperature phase, the paramagnetic solution is always present, for $T<T_d$, it disappears and is replaced by a non-trivial combination of states.      
To explicitly compute the complexity of the system, $\Sigma$, and grasping the physics behind it, several techniques have been proposed in the last decades. In the following, we will focus on the so-called \emph{real replica method} \cite{Monasson1995}. It consists in replicating $m$ times the system and coupling the different copies through an infinitesimally small parameter $\epsilon$, which will be sent to zero at the end of the computation after the thermodynamic limit. This attractive coupling naturally breaks the replica symmetry: it constrains the $m$ copies to be in the same metastable state yet remaining uncorrelated within a state.
As a consequence, the free energy can be simply written as $m$ times the individual contribution $f_\alpha$.
In this case:
\begin{equation}
    Z_m \equiv \sum \limits_{\alpha} e^{-Nm\beta f_\alpha(T)}= \int df e^{N\left[ \Sigma(f,T)-\beta m f\right]}
\end{equation}
and, by evaluating the integral by saddle-point method, we get a similar expression to (\ref{complexity_min})
\begin{equation}
\left . \frac{\partial \Sigma(f, T)}{\partial f}  \right \vert_{f^{*}(m,T)} = \beta m 
\label{slope}
\end{equation}
where one can immediately notice that the breaking parameter $m$ has a clear counterpart in the study of the complexity.
Fixing the temperature and varying only the parameter $m$, one can expand the complexity around its minimum and obtain:
\begin{equation}
    \Sigma(f,T) \approx \Sigma(f_0)+a(T)(f-f_0) + ...
\end{equation}
which implies that for $m=1$ the static transition is the solution of the equation $a(T)=\frac{1}{T}$, whereas at small $m$ the condition is precisely replaced by Eq. (\ref{slope}).

The free-energy density of a system of $m$ different copies can thus be rewritten as \footnote{To be more definite, the exact protocol requires the computation of $\phi(m,\beta)=-\frac{1}{\beta N} \lim \limits_{n \rightarrow 0} \partial_n \overline{(Z^{m})^n}$, reproducing a system of $(m \times n)$ copies and eventually taking the limit $n \rightarrow 0$.}
\begin{equation}
\begin{split}
\phi(m, \beta)= \overline{-\frac{1}{\beta N} \ln  Z_m}= & \min \limits_f \left[ \beta m f -\Sigma(f) \right]    =\\
= & \beta m f^{*}(m,\beta)-\Sigma(f^{*}(m, T))    \ .
\end{split}
\end{equation}
Generalizing the canonical definition of entropy applied to disordered systems, the complexity is thus defined as the Legendre transform of the free energy averaged over quenched disorder.

From the parametric plot of $f^{*}(m, \beta)$ and its transform $\Sigma(m, \beta)$, one can extract the information about the behavior of the configurational entropy at any temperature and, consequently, of the associated TAP states. 
Therefore, $\Sigma$ can be explicitly calculated from $\phi(m,\beta)$ or, thanks to the identity:
\begin{equation}
\phi(m,\beta)= m F^{\text{$1$RSB}}
\end{equation}
from a direct computation of the $1$RSB free energy
\begin{equation}
\begin{split}
\Sigma=&   m^2 \frac{d}{d m} \left( \beta F^\text{1RSB} \right)  =\\
=& -m^2 \frac{d}{d m} \left( \frac{1}{n} \ln { \overline{ \int d q_d d q_1 d q_0 d h \; e^{S \mathcal{A}(q_d, q_1, q_0,h)}}} \right) \ .
\end{split}
\end{equation}
We compute then the derivative of the free energy w.r.t $m$, which eventually leads to
\begin{widetext}
\begin{equation}
\Sigma= \frac{m^2 \rho^2 \sigma^2 \beta^2}{4}(q_1^2-q_0^2) +\int \mathcal{D}z \; \ln \left[\int \frac{d t_{a_B}}{\sqrt{2 \pi}}e^{-\frac{t_{a_B}}{2}}A(z,t_{a_B})^m\right]    
-m  \int  \mathcal{D} z \frac{\int \frac{d t_{a_B}}{\sqrt{2 \pi}}e^{-\frac{t_{a_B}}{2}}A(z,t_{a_B})^m \ln A(z,t_{a_B})}{\int \frac{d t_{a_B}}{\sqrt{2 \pi}} A(z,t_{a_B})^m}
\end{equation}
\end{widetext}
where we have denoted as $A(z,t_{a_B})$:
\begin{equation}
A(z,t_{a_B}) \equiv  \int d N e^{-\beta H_\text{$1$RSB}(N,z,t_{a_B})} 
\end{equation}
We quantitatively evaluate the above expression within the $1$RSB phase to determine the nature of the emerging transition. We find strictly positive values of complexity at finite temperature and fully compatible with the results previously obtained at zero temperature \cite{Biroli2018_eco}. 
We manage then to prove that the emergent $1$RSB stable phase is actually characterized by a finite complexity, \emph{i.e.} an exponential number of metastable states.
Note also that, because the replicon $1$RSB becomes negative below the Gardner transition temperature, the complexity is well-defined only in the region for which the $1$RSB can be safely applied, {\it i.e.} in the interval $[f_0,f_G]$, $f_G$ being the free-energy density at the Gardner transition.

Within this formalism, it is also possible to reproduce the complexity curves at different and fixed $m$ starting from its equilibrium value $m^{*}$. The self-condition equation $\partial F^{\text{$1$RSB}}/\partial m=0$ gives indeed the equilibrium value $m^{*}$, which corresponds to the lowest free energy density and then to a vanishing complexity. The resulting complexity curve is expected to have an increasing trend for lower values of $m$ up to a maximum point where unstable states start to appear and dominate the thermodynamics.

% \newpage 

% \bibliography{ada.bib}

%% file: app-felix.tex
%\documentclass[12pt,a4paper]{article}
%\usepackage[utf8]{inputenc}
%\usepackage{amsmath}
%\usepackage{amsfonts}
%\usepackage{amssymb}
%\usepackage{graphicx}
%
%\usepackage{xcolor}		% For coloring math
%
%%\usepackage{natbib}
%\usepackage{bibentry}	% To cite full reference
%\nobibliography*
%
%\usepackage{caption}
%\usepackage{subcaption}
%
%\author{Felix Roy}
%
%\newcommand{\us}{$\_$}
%
%
%
%
%\begin{document}
%
%
%
%
%
%\title{Demographic noise in Lotka-Volterra\\Simulations for Ada}
%
%\author{Felix Roy, October 2019}
%
%\maketitle
%\tableofcontents

\newpage

\section{Comparison theory and numerics}

\label{Comparison-Num}

\subsection{Protocol}

For comparing with the theoretical results, we sample the dynamical system presented in Equation (\ref{dynamical_eqT}). The input parameters of a sample are the system size $S$, the interaction matrix parameters $(\mu,\sigma)$, the immigration $\lambda$, the strength of the demographic noise $T$ (temperature), and the initial condition distribution $\mathbb{P}[\{N_i(0)\}_{i=1..S}]$. In order to sample one realization of the ecosystem, we perform the following steps:

\begin{enumerate}
	\item We sample the $S$-sized symmetric interaction matrix $\alpha$, from the Gaussian distribution with scaled parameters $(\mu,\sigma)$.
	\item We sample the initial conditions $N_i(t=0)$ from the distribution $\mathbb{P}[\{N_i(0)\}_{i=1..S}]$. For instance, we use a factorized uniform distribution in $[0,1]$: $$\mathbb{P}[\{N_i(0)\}_{i=1..S}] = \prod_{i=1}^S \mathbf{1}\{N_i(0) \in [0,1] \}$$ where $\mathbf{1}\{ .\}$ is the indicator function.
	\item We sample the demographic noise $\{\eta_i(t)\}_{i=1..S}^{t=0..t_{max}}$, from the white-noise distribution, with temperature $T$.
	\item Then, all three random contributions (interactions, initial conditions and demographic noise) have been dealt with. We can then integrate deterministically the system, to end up with $\{N_i(t)\}_{i=1..S}^{t=0..t_{max}}$, where $t_{max}$ is the temporal extent for the simulation.
\end{enumerate}

Actually, the above 3 and 4 points are a bit more involved: the implementation of immigration is detailed in Appendix \ref{app:mathImmigration}%\textit{Mathematical issues for immigration implementation} below% \ref{app:mathImmigration}
, and the exact numerical scheme we used is presented in Appendix \ref{app:numericalScheme}%\textit{Our implementation} below
.  But for simplicity's sake, let's focus on this framework: we fix parameters $(S, \mu, \sigma, \lambda, T)$, we sample the three random contributions, we integrate, and we obtain the species populations over time $\{N_i(t)\}_{i=1..S}^{t=0..t_{max}}$.

% References to appendices don't print out anything, because there are no numbered sections...

When we reproduce different sets of data by keeping the same parameters, but sampling different randomness, we obtain $\{N_i^r(t)\}_{i=1..S, \, r=1..N_{sample}}^{t=0..t_{max}}$.

\subsection{Observables}

In order to compare with the theory, we need to decide on the observables. So far, there are four sources of statistics in the process: the three random parts (interactions, initial conditions and demographic noise) that we labelled with $r=1..N_{sample}$, and the species themselves $i=1..S$. In the following, we will denote $\mathbb{E}[X]$ the average over all those contributions. For example:
$$\mathbb{E}[N(t)N(t')] = S^{-1} N_{sample}^{-1} \sum_{i=1}^S   \sum_{r=1}^{N_{sample}}  N_i^r(t) N_i^r(t')  $$

It can be shown \cite{benarous_convInLaw} that if the system is large enough ($S \gg 1$) and the sampling thorough enough ($N_{sample} \gg 1$), there is a convergence in law of the process. Mainly, there is a well defined limit ($S,N_{sample} \to \infty $) that we can compare with the theory. To fix ideas, we generically use $S \sim 500$ and $N_{sample} \sim 50$, and we checked there is no $(S,N_{sample})$ dependency at this scale. More precisely, in the $S \to \infty$ limit, the free energy is self-averaging, so results should typically not depend on the realization of the sampling. Here for the numerics, as $ 1 \ll S < \infty$, we still use some averaging over the samples to get cleaner data.

The theory is a thermodynamical one, so we will first assume that if we wait for a big-enough $t_{wait}$, the system will reach a time-translationnally invariant (TTI) state. For instance, the two-time correlation $C$ is a function of the time difference:
$$\forall \; t \geq t' > t_{wait}, \qquad \mathbb{E}[N(t)N(t')] = C(t, t') = C(t-t')$$

We check this numerically. The waiting-time depends on the parameters, mainly $(\sigma,T)$. However, if we lie in the replica-symmetric (RS) phase, we can always find the TTI state, for rather small waiting times $t_{wait} \sim 10^2$.

All the comparisons we will be making are in this state ($t \geq t_{wait})$, for the RS phase. We will now use a mapping between thermodynamics properties, and dynamical ones. We use the notations from Equations (\ref{Ansatz_RS}).

$$h=\mathbb{E} \left[ N(t) \right]$$
$$q_d = C(0) = \mathbb{E} \left[ N(t)^2 \right]$$
$$q_0 = \lim_{\tau \to \infty} C(\tau) = \lim_{\tau \to \infty} \mathbb{E} \left[ N(t) N(t+\tau) \right] \sim \mathbb{E} \left[ N(t) N(t_{max}) \right]$$

The \textit{lhs} is predicted by the theory, and the \textit{rhs} are numerical observables.

\subsection{Example of numerical results in the RS phase}

\begin{figure}[hbtp]
\centering
\includegraphics[width  = \linewidth]{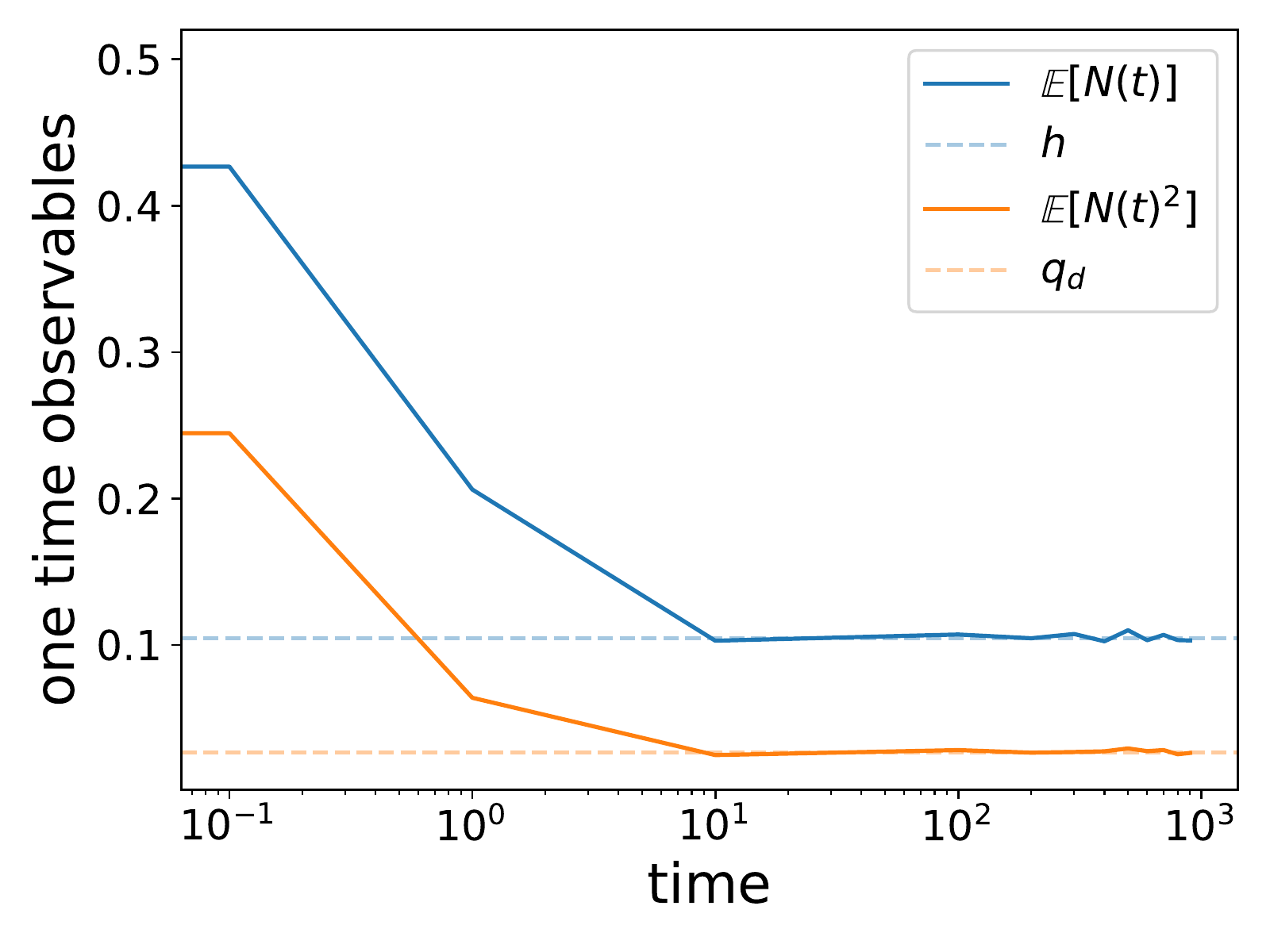}
\caption{RS one time observables converge in time. Parameters are $(S, \mu, \sigma, \lambda, T) = (500, 10, 1, 10^{-2}, 10^{-1})$. This data comes from only one sample of the process, with discrete timestep $dt=10^{-1}$. The dashed lines correspond to the read TTI value of $h$ and $q_d$.}
\label{fig:RS_example_1time}
\end{figure}

\begin{figure}[hbtp]
\centering
\includegraphics[width  = \linewidth]{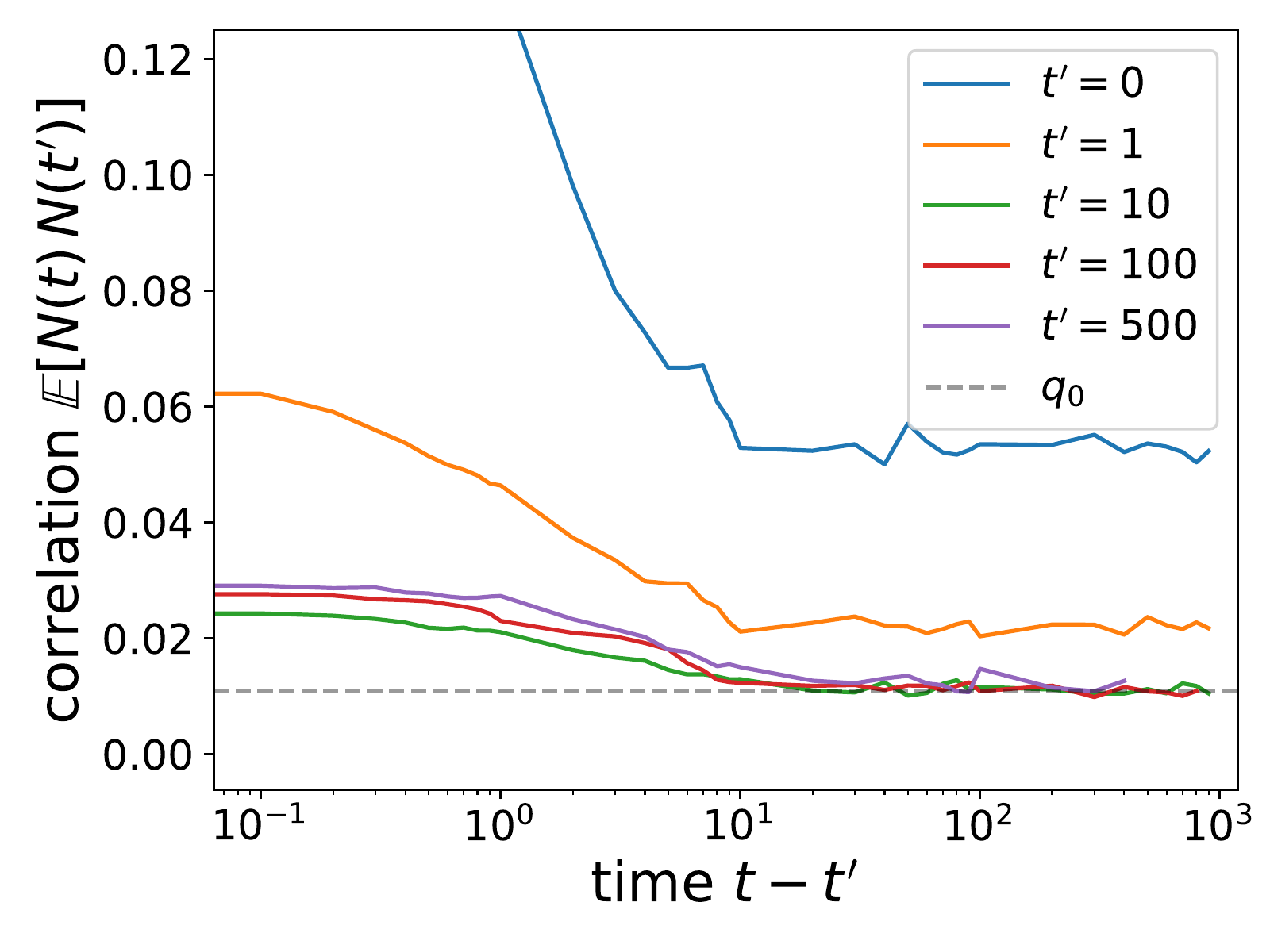}
\caption{RS correlation $\mathbb{E}[N(t)N(t')] = C(t, t')$, plotted as a function of $t-t'$ for different $t'$. This data is from the same sample as figure \ref{fig:RS_example_1time}. We see that, up to fluctuations, the correlation collapse as a function of $t-t'$ for $t' > t_{wait} \sim 20$. The dashed line correspond to the read TTI value of $q_0$. Here, the timescale for decorrelation is around $\tau_{decorrel}\sim 10$.}
\label{fig:RS_example_correlation}
\end{figure}

On figure \ref{fig:RS_example_1time}, we show that one time observables such as $\mathbb{E} \left[ N(t) \right]$ or $\mathbb{E} \left[ N(t)^2 \right]$ converge to a constant value in time. This indicates the reach of a TTI state. It can be confirmed by the collapse of two-time observables such as the correlation $\mathbb{E}[N(t)N(t')] = C(t, t')$, that we plot as $C(t-t', t')$ for different $t'$ on figure \ref{fig:RS_example_correlation}.

We can see that for $t>t_{wait} \sim 20$ here, the system is indeed TTI, at least regarding these observables. We then read the values of $h=\mathbb{E} \left[ N(t) \right]_{TTI}$ and $q_d=\mathbb{E} \left[ N(t)^2 \right]_{TTI}$ when they stabilize. And we read $q_0 = \mathbb{E} \left[ N(t) N(t_{max}) \right]_{TTI}$ on the collapse of figure \ref{fig:RS_example_correlation}.

We can also infer a relevant information from figure \ref{fig:RS_example_correlation}: the timescale for decorrelation $\tau_{decorrel}$. We are only interested in the scaling of this observable, so we introduce a rough estimate. We approximate $\tau_{decorrel}$ by the needed time so that the decorrelation decay is of $70\%$. Mathematically, $\tau_{decorrel}$ is then determined by:
$$C(\tau_{decorrel}) - C(\infty) = 0.3 \, \left( C(0) - C(\infty) \right) $$

\subsection{Match in the RS phase}
\label{matching}

\begin{figure}[hbtp]
\centering
\includegraphics[width  = \linewidth]{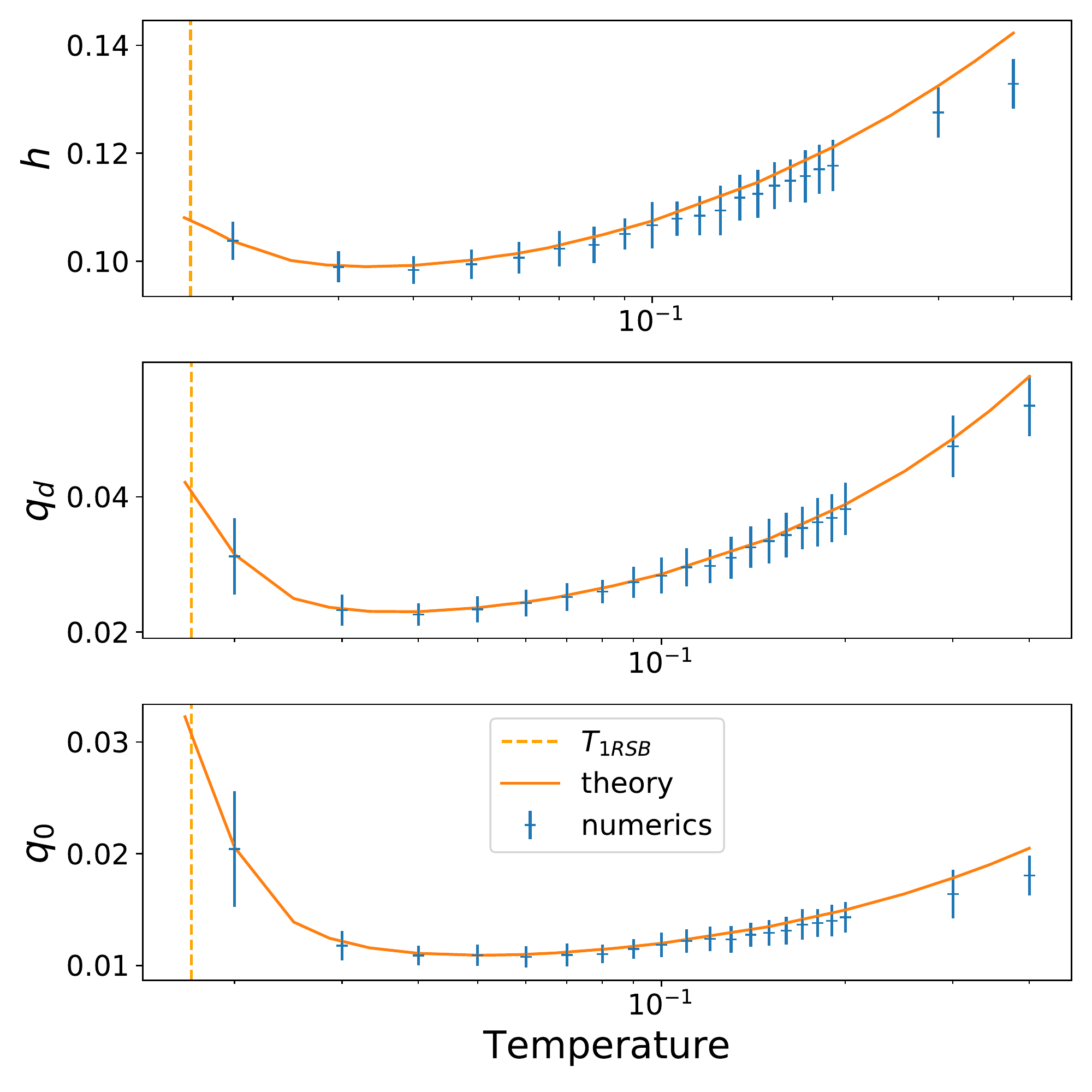}
\caption{Comparison with the theory in the RS phase. Parameters are $(S, \mu, \sigma, \lambda) = (500, 10, 1, 10^{-2})$. We consider the observables $(h,q_D,q_0)$ as a function of temperature. The orange full line is the theory predictions. Blue crosses are numerical results, error bars are taken with respects to the $N_{sample}=50$ different samples of the ecosystem. We found $t_{wait} \sim 200$ to be enough to observe TTI state in all these values of temperature, except for the last point on the left ($T=2.10^{2}$): due to slowing down of the dynamics, we had to increase the extent of the simulation and found $t_{wait} \sim 3000$. the orange dashed line correspond to the critical temperature at which the theory becomes 1RSB. Indeed, numerically we can't observe TTI state below this temperature, even increasing $t_{max}$ to $10^7$.}
\label{fig:RS_theory}
\end{figure}

The results are presented on figure \ref{fig:RS_theory}: the theory matches beautifully the numerics.

\begin{figure}[hbtp]
\centering
\includegraphics[width  = \linewidth]{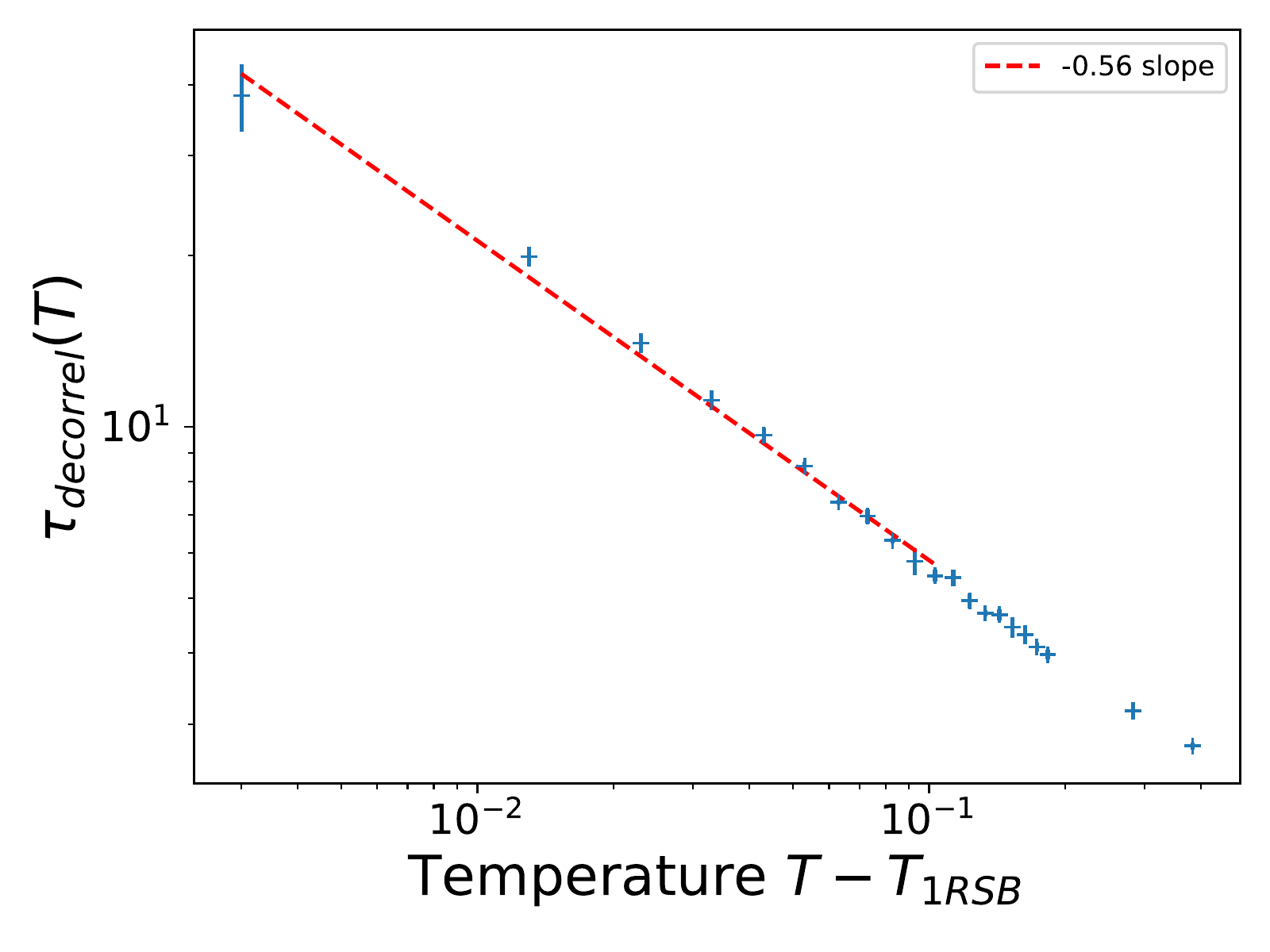}
\caption{Critical slowing down of the dynamics. We plot the decorrelation time $\tau_{decorrel}(T)$ as a function of $T-T_{1RSB}$, in a loglog scale. Blues crosses come from the same numerical data as figure \ref{fig:RS_theory}. Red dashed line is a simple fit. As we approach the transition, the system becomes slower and slower, and the dynamical timescale diverges.}
\label{fig:RS_tauDecorrel}
\end{figure}

On figure \ref{fig:RS_tauDecorrel}, we show how the timescale for decorrelation $\tau_{decorrel}$ diverges as we approach the 1RSB transition from above in temperature. Data seems to indicate a critical exponent as $\tau_{decorrel}(T) \sim \left( T - T_{1RSB} \right)^{-1/2}$.

\subsection{Rough results in the 1RSB phase}

\begin{figure}[hbtp]
\centering
\includegraphics[width  = \linewidth]{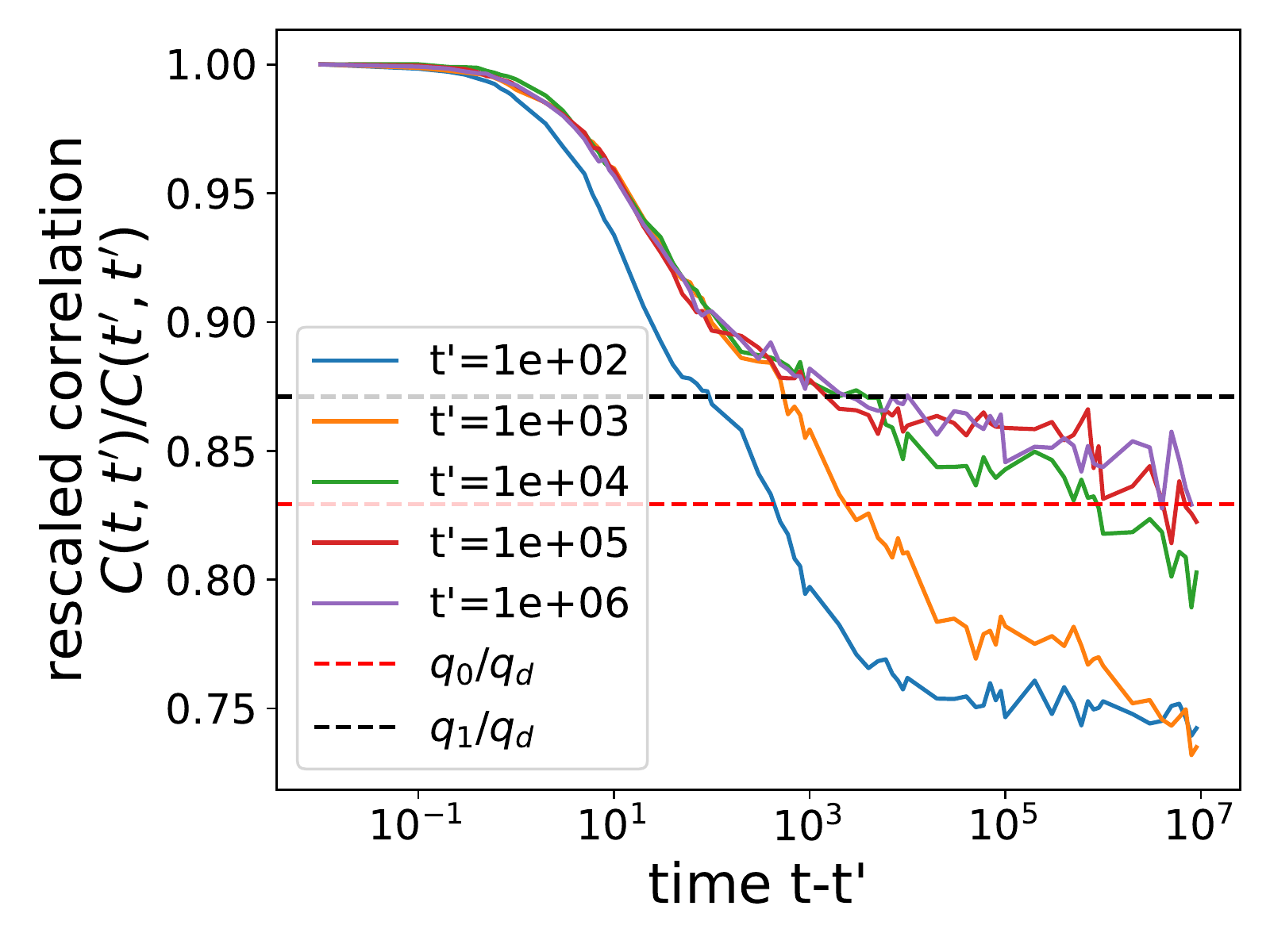}
\caption{1RSB aging. Parameters are $(S, \mu, \sigma, \lambda, T, N_{sample}) = (2000, 10, 1, 10^{-2}, 1/80, 40)$. We plot the rescaled correlation $C(t,t')/C(t',t')$ as a function of $t-t'$, for different $t'$. The different curves no longer collapse, there is no TTI state anymore. In dotted black and red lines, we respectively show the predictions for the intermerdiate and final plateau values. They do not coincide exactly with the data, but the trends correspond.}
\label{fig:1RSB_aging}
\end{figure}

In the 1RSB phase, thermodynamics indicate that the system no longer reaches a TTI state. Instead, it presents aging behaviour: the older the system is, the slower it becomes. In the simplest case of aging, there is a good understanding of the correlation decay $C(t,t')$. At equal time, the correlation is the dynamical one $C(t',t')=q_d$. Then it decorrelates quickly for $t>t'$ to an intermediate plateau $C(t,t')=q_1 < q_d$, as the system explores the neighbouring phase space. Eventually, for $t \gg t'$, the correlation decreases to a final plateau $C(t,t')=q_0 < q_1$. The timescale from the intermediate plateau to the final plateau increases with the age $t'$ of the system. These theoretical predictions are compared with numerical results on figure \ref{fig:1RSB_aging}.

\section{Mathematical issues for immigration implementation}

\label{app:mathImmigration}

In this part, we detail the mathematical issue for immigration implementation. This problem is independent on the interactions, so we drop them ($\alpha = 0$ here). We consider the following one-species Ito-stochastic process:

\begin{equation}
\frac{d N}{dt}=  N(1-N) + \sqrt{2TN} \; \eta + \lambda I(N)
\end{equation}

with white noise $\langle \eta(t) \eta(t') \rangle = \delta(t-t')$, and immigration function $I(N)$. Immigration is generically implemented so that the populations do not go too close to 0. In the usual immigration, $I(N)=1$, but we will see that this is problematic.

We want to have a hint at the stationary distribution of population $P_\infty(N)=P(N,t=\infty)$. In order to obtain it, we change variables so that the noise becomes additive, and not multiplicative any more. Here the relevant change of variables is $s(t) = \sqrt{N}(t)$, and Ito's lemma gives:

$$\frac{d s}{dt} = \frac{s^2-s^4 + \lambda I(N)}{2s} -\frac{T/4}{s} + \sqrt{2T/4} \; \eta$$

Then we use Langevin-Boltzmann to read the stationary distribution.

In the usual immigration $I(N)=1$ case, the stationary distribution is always integrable:
$$\mathcal{P}_\infty(N) = Z^{-1} N^{\lambda/T -1} \exp  \frac{1}{T} (N-\frac{N^2}{2}) $$

However, the corresponding effective potential $V_{eff}(N)$ behaves repulsively around $N=0$ only if $\lambda>T$ : 
$$ V_{eff}(N) = -N + \frac{N^2}{2} - (\lambda - T) \ln N$$

Indeed, if we introduce an approximate induced cut-off value $N_{cut}(b)$ such that $\mathcal{P}_\infty(N<N_{cut}) \sim e^{-b} \ll 1$, the scaling yields $N_{cut}(b) \sim  e^{ - b T/\lambda }$, which means that the density is still relevant up to $e^{ - b T/\lambda } \ll 1$.

Basically, this means that demographic noise with usual immigration will not prevent populations from reaching very low values. The usual immigration is not strong enough when facing demographic noise. This is indeed problematic, because whenever we will want to actually compute observables, the integrals will be dominated by the domain $N \sim 0$. This is wrong physically (important species should be the high population ones), and difficult numerically (integration is ill-defined).

In order to solve this, we can implement stronger immigration such as $I(N)=N^{-\alpha}$ with $\alpha>0$. However, another even simpler physical solution is to impose a hard repulsive boundary condition on the problem: an infinite potential at $N=\lambda$. In this case, the same steps can be performed and we obtain the stationary distribution:

$$\mathcal{P}_\infty(N) = Z^{-1} N^{-1} \exp \left[ \frac{1}{T} (N- \frac{N^2}{2}) \right]  \, \mathbf{1}\{N>\lambda\}$$

which is well-behaved. This is the solution we chose for both the theory predictions and the numerics. We will now detail in the next section how to integrate this process numerically.

\section{Numerical scheme to sample demographic noise}

\label{app:numericalScheme}

\subsection{Litterature review}

Numerical simulations need discrete time. However, when discretizing time with bounded random processes, one often encounters a non-zero probability that during one time-step the system will cross the boundary of the system (for example the $N \geq 0$ boundary in our case), and become numerically unstable. We review different solutions that have been proposed to solve this issue, and check how they deal with our Lotka-Volterra (LV) system. A more thorough review can be found in \cite{weissmann_simulation_2018}.\\

A first naive way to go around the difficulty is to change variable (sqrt, ln...). But this won't work because if the noise becomes treatable, the deterministic part becomes numerically unstable. Most articles then study the numerical integration of processes such as $\dot{N} = \alpha + \beta N + \sqrt{\sigma N} \, \eta$.\\

\cite{milstein_balanced_1998} proposes Balanced Implicit Method: they implement a clever discretization scheme so that the boundaries (positivity) are respected. The scheme amounts to Euler's for small time step. It needs a small regularization. It does not work for LV, because it needs very small regularization parameter and time-step to give good results. This is too heavy numerically.\\

\cite{pechenik_interfacial_1999} derives the exact Fokker Planck solution of a simpler system. But sampling is inefficient (rejection method). \cite{moro_numerical_2004} builds on this method by improving the sampling method, but this still isn't satisfactory. Eventually, \cite{dornic_integration_2005} improves again the method, by exactly solving (Fokker-Planck) the full process. The sampling is clever, with Poisson variables. They also indicate a way to solve more elaborate processes, which we will detail in the following section. Our strategy is heavily based on \cite{dornic_integration_2005}.

\subsection{Our implementation}

The idea from \cite{dornic_integration_2005} is to separate the process into solvable ones. More precisely, we want to solve: 

\begin{equation}
\label{eq:newScheme}
\dot{N_i} = {\color{red} \sqrt{N_i}\eta_i \;} {\color{blue} - N_i^2\;} {\color{magenta} - N_i \left( \sum \alpha_{ij} N_j -1 \right)}\qquad ... {\color{cyan}+ hardWall(\lambda)}
\end{equation}

where $hardWall(\lambda)$ implements the hard wall boundary at $N=\lambda$. We will discretize time with a timestep $dt$, and further subdivise it into three timesteps $dt' = dt/3$. We consider that only one part of the process is active during a subtimestep $dt'$. So the final scheme is the following:

\begin{enumerate}
\item From \cite{pechenik_interfacial_1999}, we know how to sample efficiently the demographic noise only
$$ {\color{red}  \tilde{N}_i(t+dt') } = \mathrm{Gamma}\left[  \mathrm{Poisson}[ \frac{ N_i(t) }{T\,dt'}]  \right] \, T\,dt' $$
This corresponds to a process which only feels the demographic noise $\dot{N_i} = \sqrt{N_i} \, \eta_i$ during $[0,dt']$. We respectively used the notation $\mathrm{Poisson}[\omega]$ ($\mathrm{Gamma}[\omega]$) for random Poisson (Gamma) variables, with parameter $\omega$.
\item Treating immigration as a reflecting wall. The particle wishes to go to $\tilde{N}_i$ but bounces on the wall.
$${\color{cyan} N_i(t+dt') = \lambda + | \tilde{N}_i(t+dt') - \lambda | }$$
\item During $[dt',2dt']$, only integrate the blue process $\dot{N_i} = -N_i^2$:
$$ {\color{blue}  N_i(t+2dt')} = \frac{N_i(t+dt')}{1+ dt' \, N_i(t+dt')}  $$
\item During $[2dt',3dt']$, only integrate the pink process $\dot{N_i} = - N_i \left( \sum \alpha_{ij} N_j -1 \right)$:
$$ {\color{magenta}  N_i(t+3dt')} = N_i(t+2dt')\exp dt' \left(1-\sum \alpha_{ij} N_j(t)\right) $$
\end{enumerate}

There are a lot of different combinations of this kind of schemes. We tried some, and chose this one after a lot of checks on simpler models for which we know the distributions at all times.

\subsection{Issues of our implementation}

After careful tests on simpler models, we used this scheme to compare with the theory. Initially we used a hardwall immigration at $\lambda = 10^{-3}$. The agreement was quite good for second degree observables ($q_d, q_0$), but not for $h$. This is due to the numerical scheme. Indeed, if $T$ is quite high ($T \gg \lambda$), the sampling of the demographic noise sends many $\mathcal{O}(1)$ species close to 0, then they bounce on the wall and end up at $N=2\lambda$. Therefore there is an induced concentration of species at $N=2\lambda$. Because of the $2\lambda$ peak, there is a subsampling of the $\mathcal{O}(1)$ populations.
	
In order to reduce this issue, we use a higher $\lambda = 10^{-2}$ in the final results that are shown on figure \ref{fig:RS_theory}. We reckon the slight discrepancy at high temperature between theory and numerics comes from this issue. A solution is still under investigation in Appendix % \textit{Improve the numerial scheme} below
\ref{app:ongoing}
. We are aware that the method is still in development. However, it is already enough at the moment to beautifully confirm the theory.

\section{Ongoing investigations}
\label{app:ongoing}

\subsection{$\chi_4$}

A cleaner numerical test for the transition RS to 1RSB would be the divergence of the $\chi_4$ correlation.
So far, we do not have enough data to present clean results, but in principle this observation should not be too difficult.

\subsection{Improve the numerical scheme}

In the current numerical scheme, we first sample pure demographic noise then implement the hard wall immigration. When doing this, a lot of trajectories do bounce on the wall, which lowers the accuracy of the scheme. A way to solve this would be to directly solve the Fokker-Planck equation associated to the whole process "demographic noise + hard wall". We reckon this can be done adapting the proof from \cite{dornic_integration_2005}.

% \newpage
% \addcontentsline{toc}{section}{References}
% \bibliographystyle{ieeetr}
% \bibliography{demoSimulation}

% \end{document}